\documentclass[manuscript=article,journal=jacsat,layout=twocolumn]{achemso}
\setkeys{acs}{articletitle = true}
\setkeys{acs}{chaptertitle = true}
\setkeys{acs}{abbreviations = false}
\setkeys{acs}{etalmode = truncate}
\setkeys{acs}{maxauthors = 10}

\usepackage[version=3]{mhchem}
\usepackage[T1]{fontenc}
\usepackage[utf8]{inputenc}
\usepackage{caption}
\usepackage{subcaption}
\usepackage{graphicx}
\usepackage[dvipsnames,table]{xcolor}
\usepackage{soul}
\usepackage{colortbl}
\usepackage{makecell}
\usepackage{multirow}
\usepackage{chemfig}
\usepackage[normalem]{ulem}
\usepackage{siunitx}
\usepackage{amsmath}
\usepackage{braket}

\usepackage{xr}
\author{Davide Faccial{\`a}}
\affiliation[CNR-IFN]{ CNR, Istituto di Fotonica e Nanotecnologie, 20133 Milano, Italy}
\alsoaffiliation{these authors contributed equally}
\author{Matteo Bonanomi}
\affiliation[DF]{Dipartimento di Fisica, Politecnico di Milano, 20133 Milano, Italy}
\alsoaffiliation{these authors contributed equally}

\author{Bruno Nunes Cabral Tenorio}
\affiliation[DTU]{Department of Chemistry, Technical University of Denmark, DK-2800 Kongens Lyngby, Denmark}
\alsoaffiliation[UAM]{Departamento de Química, Universidad Aut{\'o}noma de Madrid, Madrid 28049, Spain}

\author{Lorenzo Avaldi}
\affiliation[ISM-CNR]{CNR, Istituto di Struttura della Materia, 00133 Rome, Italy}

\author{Paola Bolognesi}
\affiliation[ISM-CNR]{CNR, Istituto di Struttura della Materia, 00133 Rome, Italy}

\author{Carlo Callegari}
\affiliation[Elettra]{Elettra-Sincrotrone Trieste S.C.p.A., in Area Science Park, 34149 Basovizza, Trieste, Italy}

\author{Marcello Coreno}
\affiliation[ISM-CNR]{CNR, Istituto di Struttura della Materia, 00133 Rome, Italy}

\author{Sonia Coriani}
\affiliation[DTU]{Department of Chemistry, Technical University of Denmark, DK-2800 Kongens Lyngby, Denmark}
\email{soco@kemi.dtu.dk}

\author{Piero Decleva}
\affiliation[UNITS]{Dipartimento di Scienze Chimiche e
Farmaceutiche, Universit{\'a} degli Studi di Trieste, I-34127
Trieste, Italy}

\author{Michele Devetta}
\affiliation[CNR-IFN]{ CNR, Istituto di Fotonica e Nanotecnologie, 20133 Milano, Italy}

\author{Na{\dj}a Do{\v s}li{\'c}}
\affiliation[RBI]{Institut Ruđer Bo\v{s}kovi{\'c}, Bijeni\v{c}ka cesta 54, 10000 Zagreb, Croatia}
\email{nadja.doslic@irb.hr}

\author{Alberto De Fanis}
\affiliation[EuXFEL]{European XFEL, Holzkoppel 4, 22869 Schenefeld, Germany}

\author{Michele Di Fraia}
\affiliation[CNR-IOM]{CNR, Istituto Officina dei Materiali, in Area Science Park, 34149 Basovizza, Trieste, Italy}
\alsoaffiliation[Elettra]{Elettra-Sincrotrone Trieste S.C.p.A., in Area Science Park, 34149 Basovizza, Trieste, Italy}

\author{Fabiano Lever}
\affiliation[POTSDAM]{Institut f{\"u}r Physik und Astronomie, Universit{\"a}t Potsdam, 14476 Potsdam, Germany}
\alsoaffiliation[DESY]{Deutsches Elektronen-Synchrotron DESY, Notkestra{\ss}e 85, D-22607 Hamburg, Germany}

\author{Tommaso Mazza}
\affiliation[EuXFEL]{European XFEL, Holzkoppel 4, 22869 Schenefeld, Germany}

\author{Michael Meyer}
\affiliation[EuXFEL]{European XFEL, Holzkoppel 4, 22869 Schenefeld, Germany}

\author{Terry Mullins}
\affiliation[EuXFEL]{European XFEL, Holzkoppel 4, 22869 Schenefeld, Germany}

\author{Yevheniy Ovcharenko}
\affiliation[EuXFEL]{European XFEL, Holzkoppel 4, 22869 Schenefeld, Germany}

\author{Nitish Pal}
\affiliation[Elettra]{Elettra-Sincrotrone Trieste S.C.p.A., in Area Science Park, 34149 Basovizza, Trieste, Italy}

\author{Maria Novella Piancastelli}
\affiliation[Sorbone]{Sorbonne Universit{\'e}, CNRS,Laboratoire de Chimie Physique-Matiere et Rayonnement, LCPMR, Paris F-75005, France}
\alsoaffiliation[Uppsala]{Department of Physics and
Astronomy, Uppsala University, Uppsala SE-75120, Sweden}

\author{Robert Richter}
\affiliation[Elettra]{Elettra-Sincrotrone Trieste S.C.p.A., in Area Science Park, 34149 Basovizza, Trieste, Italy}

\author{Daniel E. Rivas}
\affiliation[EuXFEL]{European XFEL, Holzkoppel 4, 22869 Schenefeld, Germany}

\author{Marin Sapunar}
\affiliation[RBI]{Institut Ruđer Bo\v{s}kovi{\'c}, Bijeni\v{c}ka cesta 54, 10000 Zagreb, Croatia}

\author{Bj{\"o}rn Senfftleben}
\affiliation[EuXFEL]{European XFEL, Holzkoppel 4, 22869 Schenefeld, Germany}

\author{Sergey Usenko}
\affiliation[EuXFEL]{European XFEL, Holzkoppel 4, 22869 Schenefeld, Germany}

\author{Caterina Vozzi}
\affiliation[CNR-IFN]{ CNR, Istituto di Fotonica e Nanotecnologie, 20133 Milano, Italy}

\author{Markus G{\"u}hr}
\affiliation[POTSDAM]{Institut f{\"u}r Physik und Astronomie, Universit{\"a}t Potsdam, 14476 Potsdam, Germany}
\alsoaffiliation[DESY]{Deutsches Elektronen-Synchrotron DESY, Notkestra{\ss}e 85, D-22607 Hamburg, Germany}
\alsoaffiliation[UH]{Institut für Physikalische Chemie, Fachbereich Chemie, Universität Hamburg, Martin-Luther-King-Platz 6, 20146 Hamburg, Germany}

\author{Kevin C. Prince}
\affiliation[Elettra]{Elettra-Sincrotrone Trieste S.C.p.A., in Area Science Park, 34149 Basovizza, Trieste, Italy}
\email{kevin.prince@elettra.eu}

\author{Oksana Plekan}
\affiliation[Elettra]{Elettra-Sincrotrone Trieste S.C.p.A., in Area Science Park, 34149 Basovizza, Trieste, Italy}
\email{oksana.plekan@elettra.eu}

\normalsize
\title[TR-XPS]{Unraveling the relaxation dynamics of Uracil: insights from time-resolved X-ray photoelectron spectroscopy}

\abbreviations{TR-XPS (time-resolved X-ray photoelectron spectroscopy), UV( ultra-violet), FEL (Free Electron Laser), FC (Franck-Condon), BE (Binding energy), CoIn (conical intersection), IC (internal conversion), ISC (intersystem crossing), SH (surface-hopping), RASPT2 (second-order perturbation theory restricted active space), RASSCF (restricted active space self-consistent field), FWHM (full width at half maximum), HGS (hot ground state)}


\begin{document}

%\end{document}
\cleardoublepage
%%%%%%%%%%%%%%%%%%%%%%%%%%%%%%%%%%%%%%%%%%%%%%%%%%%%%%%%%%%%%%
%%%                    ABSTRACT 
%%%%%%%%%%%%%%%%%%%%%%%%%%%%%%%%%%%%%%%%%%%%%%%%%%%%%%%%%%%%%%%%%
\begin{abstract}
We report a study of the electronic and nuclear relaxation dynamics of the photoexcited  RNA base uracil in the gas phase, using time-resolved core level photoelectron spectroscopy together with high level calculations. The dynamics was investigated by trajectory surface-hopping calculations, and the core ionization energies were calculated for geometries sampled from these. The molecule was excited by a UV laser and dynamics was probed on the oxygen, nitrogen and carbon site by core electron spectroscopy. Assuming a particular model, we find
that the initially excited $S_2(\pi\pi^*)$ state of uracil decays with a time constant of 17 ± 4 fs to the ground state directly, or to the $S_1(n\pi^*)$ state via internal conversion. We find no evidence that the $S_1(n\pi^*)$ state decays to the ground state by internal conversion; instead it decays to triplet states with a time constant of 1.6 ± 0.4 ps. 
Oscillations of the $S_1(n\pi^*)$ state O 1s intensity as a function of time correlate with those of calculated C4=O8 and C5=C6 bond lengths, which undergo a sudden expansion following the initial $\pi \to \pi^*$ excitation. We also observe oscillations in the mean energy of the main line (core ionized ionic state), which we tentatively assign to dynamics of the hot ground state. Our calculations support our interpretation of the data, and provide detailed insight into the relaxation processes of uracil.

\end{abstract}
%%%%%%%%%%%%%%%%%%%%%%%%%%%%%%%%%%%%%%%%%%%%%%%%%%%%%%
%%            INTRODUCTION 
%%%%%%%%%%%%%%%%%%%%%%%%%%%%%%%%%%%%%%%%%%%%%%%%%%%%%%
\section{Introduction}

The newly available femtosecond short wavelength pulses produced by synchrotrons, high harmonic generation (HHG), and free-electron laser (FEL) sources have enabled modern time-resolved experimental techniques to become an outstanding tool to probe ultrafast dynamical processes in nature\cite{kraus2018ultrafast}. 
In particular, time-resolved photoelectron spectroscopy (TR-PES)\cite{schuurman2022time,stolow2004femtosecond} has been recognized as an excellent method for monitoring photochemical reaction pathways in molecular systems including the non-adiabatic dynamics taking place at conical intersections (CoIns) \cite{von2018conical,polli2010conical,schuurman2018dynamics,mai2020}.

The vast majority of femtosecond TR-PES experiments with isolated molecules have been based on laboratory lasers at visible and extreme ultraviolet (EUV) wavelengths \cite{stolow2003femtosecond,fielding2018using,suzuki2012time}, and have probed the valence levels. Fewer studies are based on core level photoemission, which requires soft to hard X-rays. A key advantage of TR-PES with X-rays (TR-XPS) arises from the highly localized nature of core electrons, making them particularly sensitive to their specific chemical environment. \cite{ brausse2018time,brausse2021real,al2022observation,leitner2018time,mayer2022following,mayer2024time,gabalski2023time} In addition, the technique is quantitative to a good approximation, and the signal intensity reflects the population of a given state, which is not the case with other methods such as valence PES, and X-ray absorption.

TR-XPS with FELs has been widely applied to condensed matter, where the high density of the target and the extreme intensity of the pulses lead to space charge problems unless the intensity is strongly attenuated \cite{hellmann2010ultrafast,hellmann2009vacuum}. In the gas phase, the low density of the target reduces space charge effects, but they are still a challenge. 
 
\begin{figure*}[!htb]
\includegraphics[width=6.5in]{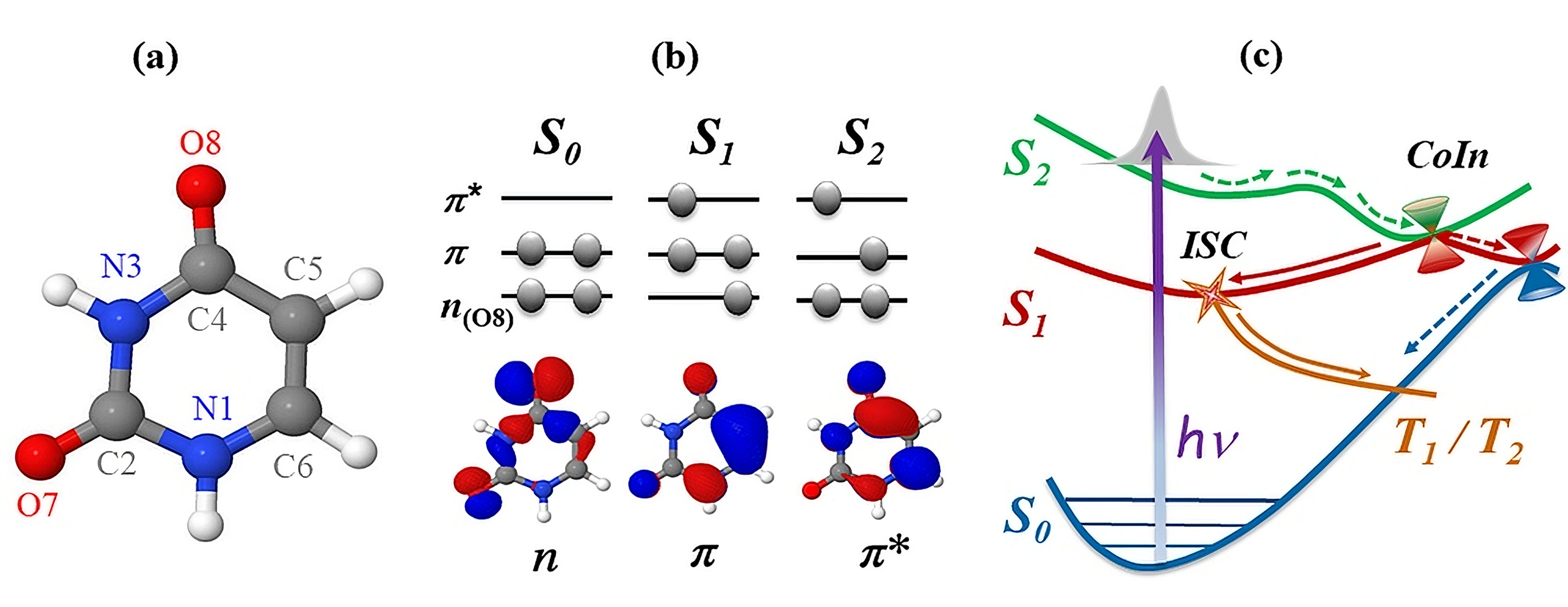}
\caption{(a) Schematic structure of uracil with atom numbering.  (b) Configurations of the ground $S_0$ and two singlet excited $S_1$, $S_2$ states of uracil together with $n$, $\pi$ and $\pi^*$ molecular orbitals. (c) Sketch of uracil relaxation upon UV excitation with potential energy
curves (PEC) of the excited singlet (green, red) and triplet (orange)
valence states along with the ground state (blue). The conical intersections (CoIn) between $S_2$/$S_1$ and $S_1$/$S_0$ PEC together with $S_1$/$T_{1,2}$ intersystem crossing (ISC) are indicated. Experimentally observed direct and indirect relaxation pathways are specified by the dashed and solid color arrows, respectively.}  
\label{fig:FIG.1} 
\end{figure*}

In this work, we applied TR-XPS with FELs to investigate the complex photoinduced dynamics of isolated uracil, one of the RNA bases (C$_4$H$_4$N$_2$O$_2$, Figure~\ref{fig:FIG.1}(a)). 
This pump-probe technique, exploiting a UV pump for the creation of excited electronic states, and a selective X-ray probe for the ionization of the C, N and O 1s core levels, allowed us to gain more insight into the ultrafast electronic relaxation processes of nucleobases.

Over the last decade, several experimental and theoretical studies have been focused on the time-resolved X-ray absorption (TR-XAS) of organic chromophores, where core level absorption, which in some respects is experimentally easier than photoelectron spectroscopy, is used as probe of photoinduced dynamics, see e.g. Refs.~\citenum{wolf2017probing,bhattacherjee2018ultrafast,wolf2019photochemical,Vidal2019,kjonstad2024photoinduced, hua2019transient,bäuml2023following}. A few theoretical studies of core phototoelectron spectra of UV-excited DNA and RNA bases have also appeared in recent years\cite{Vidal:XPS,Vidal:XPS:Erratum,mu2023identification}. Mukamel and co-authors have proposed a new TRUECARS (transient redistribution of ultrafast electronic coherences in attosecond Raman signals) technique\cite{keefer2020visualizing} and have simulated the high temporal resolution TR-PES signal in order to monitor the photorelaxation of uracil passing through a CoIn \cite{cavaletto2022electronic}.
However, studies of TR-XPS of gaseous nucleobases remain rather limited.\cite{Vidal:XPS:Erratum,Vidal:XPS,mayer2022following,mayer2024time,mu2023identification}

A combined experimental and theoretical study of the ground state ($S_0$ or GS) XPS spectra of uracil has been previously published \cite{feyer2009tautomerism}. New calculations for vibrationally resolved XPS of common biomolecules have also been recently reported\cite{wei2024}. The photoionization dynamics of uracil for both the valence and inner shells were characterized in the framework of density functional theory (DFT) approach.\cite{Toffoli2007DensityFT}

Figure~\ref{fig:FIG.1}(b) displays the Highest Occupied Molecular Orbital (HOMO), denoted $\pi$, and the non-bonding HOMO-1 orbital, denoted $n$, along with the Lowest Unoccupied Molecular Orbital (LUMO), $\pi^*$. At the Franck-Condon (FC) geometry, the lowest energy excited state $S_1$  has $n\pi^*$ character and arises from the excitation of a lone pair electron initially localized on the O8 atom (see Figure~\ref{fig:FIG.1}(a, b)) to the delocalized $\pi^*$ orbital. 
Note that the $S_1$ state is optically dark, i.e., not accessible from the $S_0$ state {\it{via}} one-photon absorption \cite{fedotov2021excited}. 
The second excited state $S_2$ is optically bright and corresponds to a $\pi\pi^*$ transition, with an electron promoted from the bonding $\pi$ orbital (localized mostly on the C5=C6 and C4=O8 bonds) to the corresponding antibonding $\pi^*$ orbital (see Figure~\ref{fig:FIG.1}(b)). 

There is a significant difference in charge distribution between the $\pi\pi^*$  and $n\pi^*$  states, particularly at the O8 oxygen atom (see Figure 1(b)). 
Hence, these excited states should have distinctly different spectra, and observable changes are expected in the O8 1s photoelectron signal. Recently, calculations by Vidal et al. \cite{Vidal:XPS,Vidal:XPS:Erratum} predicted clear differences between the theoretical XPS spectra at the oxygen K-edge of the two exited states, and at the ground state of uracil. In particular, a chemical shift of about 4~eV from the GS was predicted for the O8 core ionization of the $S_1$ state.
For the N 1s and C 1s core electrons, the changes in local charge density are smaller, so that the shifts were less pronounced, but still expected to be observable in TR-XPS spectra \cite{Vidal:XPS:Erratum,Vidal:XPS}.

The photodynamics of uracil has been investigated both experimentally and theoretically and we refer to Ref.\citenum{chakraborty2021time} for a recent summary. These studies have found that upon excitation to the $S_2(\pi\pi^*)$ state, the relaxation of uracil back to $S_0$ takes place non-radiatively, along two, possibly three, competing pathways depicted in Figure~\ref{fig:FIG.1}(c). This includes a direct relaxation along the $\pi\pi^*$ state, in which at the $S_2$/$S_1$ CoIn the $S_1$ population remains in the $\pi\pi^*$
state, and reaches the CoIn with the $S_0$ state (Figure~\ref{fig:FIG.1}(c), green/red/blue dashed arrows) \cite{chakraborty2021time,matsika2004radiationless,richter2014ultrafast,chakraborty2021faraday,milovanovic2021simulation}.
We remind the reader that the label $S_2$ and $S_1$ refers to the energetic order of the (adiabatic) electronic states, while the orbital notation $\pi\pi^*$ and $n\pi^*$ describes the electronic character of these states, that is, the diabatic states whose population is monitored in the experiment.

In the indirect pathway, the population switches from the $\pi\pi^*$ to the $n\pi^*$  state at the $S_2$/$S_1$ CoIn (Figure~\ref{fig:FIG.1}(c), red solid arrow). In the $S_1(n\pi^*)$  state the population remains trapped for several picoseconds and eventually decays by intersystem crossing (ISC) to the triplet manifold ($T_1/T_2$) (Figure~\ref{fig:FIG.1}(c), orange solid arrow).\cite{richter2014ultrafast,etinski2009intersystem,yu2016internal} Some early works have also suggested trapping of the population in the $S_2(\pi\pi^*)$ state and decay {\it{via}} ring opening \cite{nachtigallova2011nonadiabatic}. Within this general framework, the time constants associated with specific relaxation steps are still under debate \cite{serrano2007intrinsic,climent2007intrinsic,etinski2009intersystem,karak2023photophysics,ullrich2004electronic,yu2016internal,richter2014ultrafast,matsika2004radiationless, chakraborty2021time, hudock2007ab,lan2009photoinduced,nachtigallova2011nonadiabatic,milovanovic2021simulation,chakraborty2021faraday}.
However, a recent TR-PES experimental study of gaseous uracil with unprecedented time resolution has demonstrated the return of a fraction of the excited state population to the ground state, in $\approx$ 17 fs \cite{miura2023formation,karashima2025exploring}.

This paper presents a detailed study of the O 1s core-level photoelectron spectra of gaseous uracil as a function of the pump-probe delay, together with preliminary data for N and C 1s, to monitor the relaxation pathway of uracil after UV excitation. We exploit the fact that core binding energies provide quantitative local chemical information about the probed atoms in a given system. To obtain a detailed insight into both the mechanism and timescale of ultrafast processes in isolated uracil, we have coupled extensive non-adiabatic dynamics simulations with high-level multireference calculations of TR-XPS signals. We determine time constants for key processes, and we observe the effects of nuclear dynamics on the XPS signals.
%------------------------
%       METHODS
%-----------------------
\section{Methods}
\subsection{Experimental Section}
The experiment was performed at the Small Quantum Systems (SQS) instrument located at the European X-ray Free-Electron Laser (XFEL) facility~\cite{Decking_2020,Mazza:yi5133,tschentscher2017photon}. 
Isolated uracil molecules were irradiated by femtosecond soft X-ray pulses at the Atomic-like Quantum Systems (AQS) experimental station. A photon energy of 600 eV was chosen, which is sufficient to ionize all three O, N and C 1s core levels of uracil (see Supporting information (SI) for the C 1s spectra and further details). Uracil was evaporated by a capillary oven at a temperature of 433 K into an ultrahigh vacuum chamber, with a sample density of about 10$^{12}$ cm$^{-3}$ in the interaction region.\cite{McFarland_2014} 

The SASE3 soft X-ray undulator was tuned to provide X-ray pulses centered at a photon energy of 600 eV with a 5 eV  full width at half maximum (FWHM) bandwidth and mean pulse energy of 6.8 mJ. 
The FEL pulses passed through a gas attenuator reducing their energy to \text{238 }$\mu$J \cite{sinn}. 
The focus of the X-ray beam was set to a diameter of approximately \text{100} $\mu$m (FWHM) to minimize non-linear effects. 
The FEL pulses had an estimated duration of approximately 30 fs as inferred from the electron bunch charge of 250 pC \cite{serkez2021wigner, serkez2018rosa}. 

The beamline monochromator was used to reduce the FEL bandwidth to 0.136 eV (FWHM), which corresponds to a resolving power of E/$\Delta$E=4.4$\times$\text{10}$^{3}$ and resulting in a pulse energy of \text{0.21} $\mu$J on the target 
(see Figure~S1, section~S1, SI).
%(see Figure~\ref{SI-fig:FIG.1}, section~\ref{SI-section:1}, SI).

The monochromator energy scale was calibrated using the well-known atomic Ne 1s–3p resonance at 
867.12 eV~\cite{Gerasimova:ok5074} 
(see Figure S2, section S1, SI).
%(see Figure~\ref{SI-fig:FIG.2}, section~\ref{SI-section:1}, SI).

In the experimental ground state O 1s spectrum\cite{feyer2009tautomerism}, the two non-equivalent oxygen core levels were not resolved, and appeared as a single peak with a maximum at 537.6 eV. 
We applied a fine calibration (0.1 - 0.2 eV) to the energy scale using the literature value to account for possible drifts of the monochromator or spectrometer. The offset was determined from ground state spectra and applied to them and to the following excited state spectra.

Uracil molecules were excited into the bright $S_2$ electronic state by the third harmonic of an optical laser operating at $\approx$ 800 nm \cite{Pergament:16} and synchronized to the X-ray pulses. The central wavelength of the pump pulses was 264 nm (4.70 eV), with  pulse energy around 5 $\mu$J and a duration of 75 fs. The choice of this pulse energy was made based on the photoion time-of-flight (TOF) mass spectra, taken in order to check that uracil was not excessively pumped by the UV laser 
(see Figure S3, section S1, SI). 
%(see Figure~\ref{SI-fig:FIG.3}, section~\ref{SI-section:1}, SI).
The duration of the UV pulses was measured by cross-correlation measurement\cite{grychtol2021timing} 
(see Figure S4, section S1, SI) 
%(see Figure~\ref{SI-fig:FIG.4}, section~\ref{SI-section:1}, SI) 
and the focus diameter was 150 $\mu$m (FWHM).

The initial temporal and spatial overlap between UV and soft X-ray pulses was established by using a higher pump energy and observing the depletion of the normal Auger signal of uracil. The depletion is induced by fragmentation of the molecule by the UV laser which leads to photoproducts with different Auger spectra. Subsequently, a more precise temporal overlap condition was found and routinely checked by monitoring the formation of laser-induced sidebands around the 1s photoline.

A delay stage was used in order to vary the delay between the {UV} pulses and the {X-ray} pulses. A nominal step size of 20 fs was used in the range from $-$200 fs to +500 fs, where a negative delay indicates that the X-ray pulse arrived before the UV pulse. 
In addition, spectra were also acquired at fixed delays of $\SI{1}{ps}$, $\SI{10}{ps}$, $\SI{100}{ps}$ and $\SI{1}{ns}$, to capture the slower dynamics. 
The nominal delays were corrected by means of the pulse arrival-time monitor \cite{Viti:ICALEPCS2017-TUPHA125} and rebinned. A bin size of 18 fs was chosen so that all bins have similar statistics.

Photoelectron spectra as a function of the time delay between the UV and soft X-ray pulses were recorded with the magnetic bottle electron spectrometer (MBES) of the experimental station. The X-ray photons were linearly polarised perpendicular to the axis of the magnetic bottle spectrometer and the UV polarisation. In order to enhance the energy resolution of the MBES spectrometer, 1s electrons from oxygen, nitrogen, and carbon were slowed down to $\sim 25$ eV final kinetic energy using appropriate retardation voltages. The MBES spectrometer was calibrated by measuring Ne 1s spectra while varying the photon energy. The MBES resolution measured at the nominal retardation of 45~V was estimated to be E/$\Delta$E $\approx$ 0.03 
(see Figure S5, section S1, SI).
%(see Figure~\ref{SI-fig:FIG.5}, section~\ref{SI-section:1}, SI).  
The time-of-flight spectra were converted to an energy scale and binned at intervals of 100 meV.

Photoelectron spectra of the ground and photoexcited states of uracil were measured in consecutive shots by firing the UV laser at half the repetition frequency of the XFEL: with the UV laser off, the ground state was measured; with the UV laser on, the excited state plus a fraction of ground state molecules were measured. 
If the fraction of molecules excited is denoted $f$, then the fraction of ground state molecules in this case is $1-f$. The raw data were analyzed using two approaches. Firstly, the UV-off spectra were subtracted directly from the UV-on spectra, yielding difference spectra in which the excited state features were positive, and the ground state features were negative, that is, they appeared as depletion of the signal. Secondly, we performed a scaled subtraction for the fraction of ground-state molecules in the UV-on spectrum\cite{warne2021}, thereby eliminating negative features. The fraction of excited molecules $f$ was estimated using the procedure described in the Supporting Information (see section S2).

The integral of the O and N 1s core level signal of the photoexcited sample, including the depleted ground state signal, was compared to the integral of the ground state spectrum and was found to be equal within experimental error. This rules out an excessive pump pulse energy which may have led to 2-photon ionization, producing artifacts in the
experimental spectrum. Core ionization of valence ionized  molecules produces photoelectrons whose energy is shifted outside the measured energy window (ionized molecules have higher ionization potential than the corresponding neutral molecule), so that, in the presence of strong valence ionization by the pump laser, a decrease in the overall signal of the photoexcited sample is expected. Since the difference of the integrated signals of the ground and excited states was zero, we conclude that the fraction of the sample ionized by the pump laser only is negligible compared with the neutral photoexcited fraction.

%------------------------
%       THEORY
%------------------------

\subsection{Theoretical Section}

The computational protocol for simulating time-resolved XPS spectra is based on a trajectory surface-hopping description (SH) of the excited-state dynamics, and on a calculation of the core ionization energies and intensities from geometries sampled from SH trajectories. The accuracy of this protocol has been illustrated in recent publications.\cite{pitesa2021,travnikova2022photochemical} SH calculations restricted to singlet electronic states were performed by \citet{milovanovic2021simulation} using the SCS-ADC(2) method\cite{schirmer1982adc}\cite{dreuw2015adc} and the aug-cc-pVDZ basis set with an in-house code\cite{Sapunar2019} linked to the Turbomole program package\cite{TURBOMOLE}.
To allow the possibility of crossing to the triplet manifold, calculations were repeated with the SHARC 3.0 program.\cite{SHARC3.0,May2015}. The required spin-orbit matrix elements were computed using the spin-orbit mean field (SOMF) method implemented in the Orca 5.0 program package.\cite{Orca5.0}

Initial geometries and momenta were selected from 1000 randomly sampled geometries according to their oscillator strengths in the 4.7-4.8 eV excitation window. In all cases, the calculations were initiated in the $S_2(\pi\pi^*)$  state. 
Newton's equations for nuclear motion were integrated for 1000 fs with time steps of $0.5$ fs, using the velocity-Verlet algorithm. The local diabatization formalism was used to propagate the electronic wave function and compute the hopping probabilities.\cite{Granucci2001}. 
The energy-based decoherence procedure of \citet{Granucci2007} with $\alpha$  = 0.1$E_h$ was used.

Time-resolved XPS spectra of uracil were computed at early pump-probe delay times using 48 SH trajectories 
(see section S4, SI), 
%(see section~\ref{SI-section:4}, SI), 
and employing the RASPT2 method combined with the aug-cc-pVDZ basis set. The active space utilized in the restricted active space self-consistent field (RASSCF) calculations is divided into three segments: RAS1, RAS2, and RAS3.\cite{RASPT2-1,RASPT2-2} 
RAS1 comprises the relevant core orbitals, RAS2 includes seven valence-occupied orbitals, and RAS3 is formed by two $\pi^*$ orbitals, each capable of accommodating a maximum of two electrons.

Core-hole states were calculated by enforcing a single hole in RAS1 using the HEXS projection technique\cite{HEXS-RAS}, available in {\sc OpenMolcas}\cite{Openmolcas-JCTC}. RASSCF orbitals of initial valence-excited and final core-ionized states were obtained by state averaging over 10 and 20 states, respectively. The state-averaged active orbitals, computed at the equilibrium geometry of the ground state, are schematically represented in Figure 
S11 (see section S4, SI). 
%\ref{SI-fig:FIG.11} 
%(see section~\ref{SI-section:4}, SI). 

To incorporate dynamical correlation effects, the extended multi-state restricted-active-space perturbation theory of the second order (XMS-RASPT2) approach \cite{XMS-CASPT2} was employed in the reference space. An imaginary level shift of 0.35$E_h$ was applied to avoid intruder-state singularities.

The simulated XPS spectra for both the ground state and valence excited states of uracil were generated through convolution of the computed ionization energies and the squared norms of Dyson orbitals, that is, pole strengths, employing a Lorentzian function of typically 0.4 eV (FWHM). Dyson orbitals were computed following the procedure outlined in Ref.~\citenum{Tenorio:Molecules:2022}, utilizing the RASSI\cite{RASSI-MALMQVIST} module of OpenMolcas\cite{Openmolcas-JCTC} with the RASPT2 energies and the perturbatively modified (mixed) RASSCF transition densities.\cite{BATTAGLIA2023135}
To assess the quality of the XPS spectra derived from the squared norms of the Dyson orbitals, we also calculated cross-sections using an explicit description of the electronic continuum obtained at the DFT level with a linear combination of atomic orbitals (LCAO) B-spline basis, employing the Tiresia code \cite{TOFFOLI2024109038}. A detailed comparison is provided in
Figure S12 (see section S5, SI).
%Figure~\ref{SI-FIG.12} (see section~\ref{SI-section:5}, SI).

%%%%%%%%%%%%%%%%%%%%%%%%%%%%%%%%%%%%%%%%%%%%%%%%%%%%%%%%%%%%%%%%%%%%%
%% 
%%        RESULTS AND DISCUSSION 
%%
%%%%%%%%%%%%%%%%%%%%%%%%%%%%%%%%%%%%%%%%%%%%%%%%%%%%%%%%%%%%%%%%%%%%%
\section{Results}
\subsection{Oxygen K-edge} 
\begin{figure*}[!ht]
\includegraphics[width = 5 in]{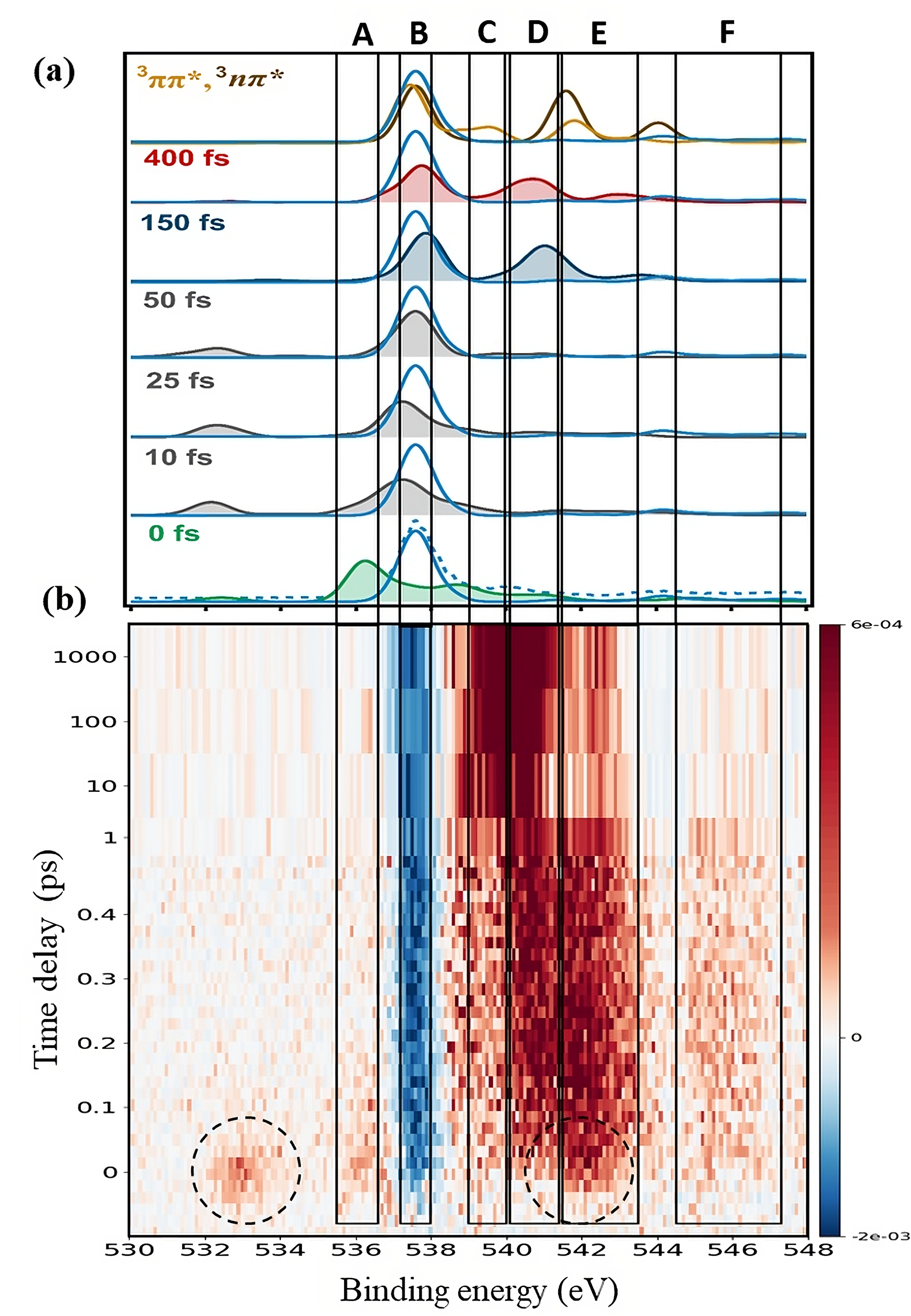}
\caption{(a) Calculated uracil O 1s spectra of the ground state (light blue solid line), singly excited states (green, grey, dark blue and red lines, with shading) at selected time delays 
%(see Table~\ref{SI-tab:trajector}, section~\ref{SI-section:4}, SI), 
(see Table S2, Section S4, SI), 
and triplet ${}^3\pi\pi^*$ (orange line) and ${}^3n\pi^*$ (brown line) states. 
Theoretical spectra are shifted by 2.4 eV to lower binding energy. 
The blue dashed line represents the experimental GS spectrum. 
(b) Two-dimensional false color map of the O 1s difference spectra (UV-on minus UV-off) as a function of binding energy and the time delay (red: positive signal; blue: negative signal). 
Black dashed circles indicate sidebands. 
Energy ranges (eV): (A) 535.9 - 536.6; (B) 537.2 - 538.0; (C) 539.0 - 540.0; (D) 540.1 - 541.5; (E) 541.6 - 543.5; (F) 544.5 - 547.5.}
\label{fig:FIG.2} 
\end{figure*}

The calculated and experimental time-resolved O 1s spectra of valence-excited uracil are shown in Figures~\ref{fig:FIG.2}(a) and~\ref{fig:FIG.2}(b), respectively. 
Table~\ref{tab:exO1s} 
summarizes the theoretical 
O~1s binding energies (BEs) for the $S_0$, $S_1$, and $S_2$ states at the equilibrium geometry of the ground state, and their corresponding ionization characters. 

The present computed energies for the two O 1s electrons in the GS are 539.95 and 540.05 eV, about 2.4 eV higher than the experimental value, so all theoretical spectra in %Figure 2(a) 
Figure~\ref{fig:FIG.2}(a)  
have been rigidly shifted by $-2.4$~eV. 
For the $S_2$ state, the calculated binding energy of O8 with final state configuration O8 $1s^{-1} \pi\pi^*$ was 538.3 eV. The weak feature computed near 534 eV is assigned to a shake-down transition of the $S_2$ state.\cite{moitra2021inner} 
Note that the theoretical shake-down energy does not match exactly the experimental excitation energy.
For the $S_1$ state, the BE of O8 1s with the final state configuration O8 $1s^{-1} n\pi^*$, was calculated to be 4 eV higher than the GS peak (see 
Table~\ref{tab:exO1s}), 
consistent with the previous prediction of Vidal et al.~\cite{Vidal:XPS,Vidal:XPS:Erratum} In contrast to the calculated BEs of the O8 atom, those of O7 are shifted only slightly from the ground state BE in the $S_2(\pi\pi^*)$ or $S_1(n\pi^*)$ state (see Table~\ref{tab:exO1s}).

\begin{table}[!ht]
\caption{Theoretical O 1s ionization energies (in eV) and orbital character computed at the RASPT2/aug-cc-pVDZ level at the FC geometry for three initial states. Due to the unpaired electrons in the $S_1$ and $S_2$ states, primary core ionization yields final doublet states with 3 unpaired electrons, which are split into several  spin multiplets; labels L and H indicate lower and higher energy states. For a discussion of spin coupling in the case of x-ray absorption of doublet molecular cations, see, e.g., Refs.~\citenum{couto2020,lindblad2020x,vidal2020interplay,epshtein2023signatures}.}

\centering
\begin{tabular}{l|cc}
\hline\hline
 Initial state & BE (eV)  &  Ionization character  \\
\hline \hline
GS &  539.95 &  O8 $1s^{-1}$  \\
   &  540.05 &  O7 $1s^{-1}$  \\
\hline
                & 539.86 & O7 $1s^{-1} n\pi^*$ \\
$ S_1 (n\pi^*)$ & 544.02 & O8 $1s^{-1} n\pi^*$  (L)\\
                & 546.40 & O8 $1s^{-1} n\pi^*$ (H) \\
\hline
                  & 534.20 & O8 $1s^{-1}$(shake-down)\\
$ S_2 (\pi\pi^*)$ & 538.26 & O8 $1s^{-1} \pi\pi^*$ \\
                  & 539.17 & O7 $1s^{-1} \pi\pi^*$ (L)\\
                  & 540.41 & O7 $1s^{-1} \pi\pi^*$ (H)\\
\hline \hline 
\end{tabular}
\label{tab:exO1s}
\end{table}
\vspace{4mm}
\noindent
The experimental and theoretical O 1s spectra presented in Figure~\ref{fig:FIG.2} were divided into six BE ranges (A-F) in which signal intensity variations are evident. 
As noted above, data were acquired in two different time-delay intervals (see Figure~\ref{fig:FIG.2}). The first range (from $-$100 fs to 450 fs) maps ultrafast dynamical changes, while the second range (from 1 ps to 1 ns)  follows the long-lived excited states.

Difference spectra are plotted as a false-color, two-dimensional map in Figure~\ref{fig:FIG.2}(b), as a function of binding energy and time delay.
This representation shows the sum of valence excited states as positive features while negative features are due to the depletion of the ground state.
The two experimental features at 532.9 eV and 542.3 eV observed around time delay = 0 (see Figure~\ref{fig:FIG.2}(b), dashed circles) are due to the lower and upper sidebands (SBs) resulting from the simultaneous absorption of a soft X-ray photon and absorption (or emission) of a UV photon. 
The maximum intensity of the low energy sideband is equal to about 0.23 times the asymptotic depletion of the main peak. Since the SB signal occurs only when the pump and probe pulses overlap temporally, it serves as a convenient monitor for the precise measurement of their cross-correlation. 

The weak intensity observed around 536 eV in range A is assigned to the excitation of the $S_2(\pi\pi^*)$ state. 
This feature disappears in less than 50 fs, in good agreement with our dynamics calculations (see Figure~\ref{fig:FIG.2}(a)). The weakness of the $S_2$ state signal was discussed by Miura et al.,\cite{miura2023formation} using valence spectroscopy, and it is primarily due to the short lifetime of this state, which is consistent with our experimental observation. 
The fast decay of the $S_2$ state signal from range A coincides with the appearance of signal in ranges C, D, E, and F. Hence, the intensity in ranges C, D, and E is assigned to the $n\pi^*$ state, populated by $S_2 \to S_1$ internal conversion (IC). 
Furthermore, according to our calculations (see Table 1), the $S_1$ state feature exhibits an asymmetric doublet structure with an energy difference of approximately 2.4 eV 
(see Figure S12, section S5, SI) 
%(see Figure~\ref{SI-FIG.12}, section~\ref{SI-section:5}, SI) 
attributed to the spin coupling of the three unpaired electrons: one in the partially vacant \textit{n} orbital, one in the $\pi^*$ orbital and one in the 1s(O8) orbital, resulting in a final core ionized state of doublet spin multiplicity.

In range B (537.2-538 eV), the depletion of the ground state is clearly visible as a single negative peak due to ionization in the region of O7 and O8 binding energies. Our calculations indicate that the depletion is due to a large energy shift of O8 in the excited states, while the ionization energy of O7 does not change significantly 
(see Table~\ref{tab:exO1s}). This implies that the integral of the depletion signal is equal to the integral of the O8 1s signal of the excited state. Our calculations suggest that, within the first 30 fs, a portion of the excited-state population returns to the ground state, giving rise to a vibrationally hot ground state (HGS). 
This HGS has the electronic configuration of the ground state and thermal energy equal to that of the UV photon. From theory we know that the HGS signal appears slightly above the ground state energy 
%(see Figure~\ref{SI-fig:FIG.13}, section~\ref{SI-section:6}, SI). 
(see Figure S13, section S6, SI). 
These calculations show that the HGS peak shape varies with time delay up to 1 ps and is asymmetric.

In the present O 1s spectra, the core ionization of the excited states does not correspond solely to a single electron ionization from a closed-shell configuration, because there are two singly occupied valence levels, leading to a final state with three unpaired electrons. This facilitates simultaneous ionization and excitation processes, that is, 2-hole 1-particle (2h1p) transitions.\cite{moitra2021inner}
The observed enhancement between 544.5-547.5 eV (Figure~\ref{fig:FIG.2}(b), range F) is attributed to a shake-up process of the molecules in the $S_1$ state, terminating in less than 10 ps, which aligns with the gradual depopulation of the dark $S_1$ state of uracil. For example, in 
Figure~S14
(see section~S7, SI), 
%Figure~\ref{SI-fig:FIG.14} 
%(see section~\ref{SI-section:7}, SI), 
the shake-up signal appears above the 546 eV region in the spectra for a selected SH trajectory which undergoes IC to the $S_1(n\pi^*)$ state. Hence, the UV-excitation of uracil made it possible to experimentally observe final states that are only weakly excited in XPS of the ground state but have much higher relative intensity in the current O 1s TR-XPS spectra. 

\subsection{Nitrogen K-edge}
\begin{figure*}[!ht]
\includegraphics[width =5 in]{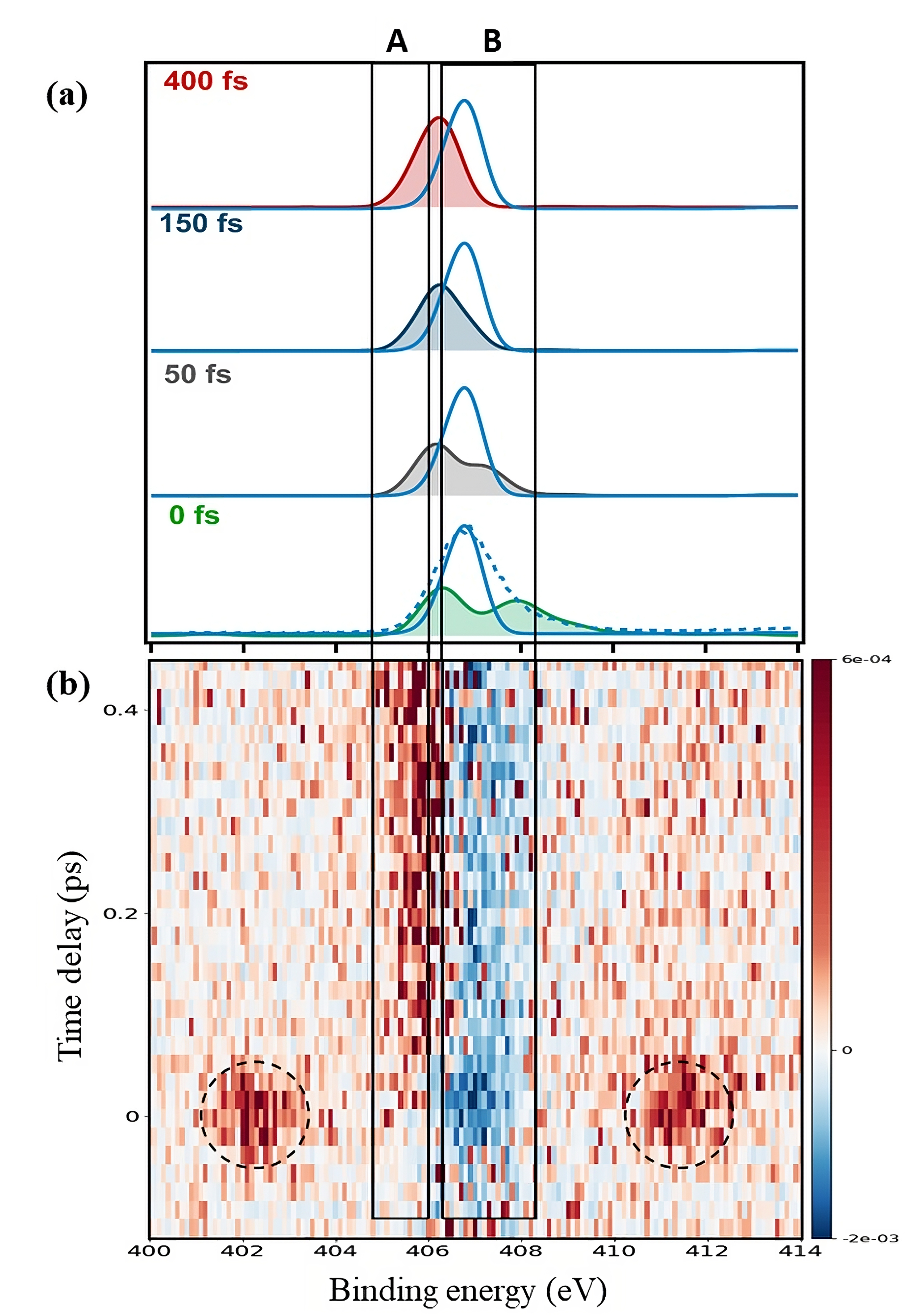}
\caption{(a) Calculated N 1s spectra of uracil in the ground (blue solid line) and singly excited states (green, grey, dark blue and red lines, with shading)  at selected time delays 
%(see Table~\ref{SI-tab:trajector}, section~\ref{SI-section:4}, SI). 
(see Table~S2, section~S4, SI).
Theoretical spectra are shifted by 2.1 eV to lower binding energy. The blue dashed line represents the experimentally measured GS spectrum.
(b) Two-dimensional false color map of the N 1s difference spectra (UV-on minus UV-off) as a function of binding energy and time delay (red: positive signal, blue: negative signal). 
Black dashed circles indicate sidebands. 
Energy ranges (eV): (A) 404.8 - 406.0 and (B) 406.3 - 408.3.}
\label{fig:FIG.3} 
\end{figure*}

The theoretical and experimental N 1s photoelectron spectra of photoexcited uracil are presented in Figures~\ref{fig:FIG.3}(a) and~\ref{fig:FIG.3}(b), respectively. 
The experimental spectra have been collected only for a short range, from $-$150 fs to +450 fs, with limited statistics, but are still sufficient to follow the charge dynamics around the nitrogen atoms.

As in the O 1s spectra, the ground state is depleted on excitation, but in contrast to the O 1s spectra, the depletion partially recovers after about 50 fs. Positive and negative sidebands are also observed, Figure~\ref{fig:FIG.3}(b), black circles. 
The ratio of the low energy sideband intensity to the asymptotic attenuation is 1.3, considerably more than in the case of oxygen.

\begin{table}[!ht]
\caption{
Theoretical N 1s ionization energies (in eV) and orbital character computed at the RASPT2/aug-cc-pVDZ level, and at the FC geometry, for three initial states.
}
\centering
\begin{tabular}{l|cc}
\hline\hline
Initial state & BE(eV)  &  Ionization character  \\
\hline \hline
GS &  408.77 &  N3 $1s^{-1}$  \\
   &  409.10 &  N1 $1s^{-1}$  \\
\hline
$ S_1 (n\pi^*)$ & 408.58 & N3 $1s^{-1} n\pi^*$ \\
                & 408.93 & N1 $1s^{-1} n\pi^*$ \\
\hline
$ S_2 (\pi\pi^*)$ & 408.38 & N3 $1s^{-1} \pi\pi^*$ \\
                  & 410.24 & N1 $1s^{-1} \pi\pi^*$ \\
\hline \hline 
\end{tabular}\\
\label{tab:exN1s}
\end{table}

The sideband intensity scales with the kinetic energy of the electron and the kinetic energy of the N 1s electron is higher than that of O 1s. Thus, the recovery of the depletion signal when the sideband channel closes is more evident for N 1s. To check this hypothesis, we show in 
%Figure~\ref{SI-fig:FIG.15} (see section~\ref{SI-section:8}, SI) 
Figure S15 (see section S8, SI) 
that by adding the intensity of the sidebands to the depletion signal, the trend of the depletion is qualitatively similar to that of oxygen.

The calculated N 1s BEs for $S_0$, $S_1$, and $S_2$ states at the equilibrium geometry of the ground state, and their corresponding ionization characters, are summarized in Table~\ref{tab:exN1s}. 
The spectrum of the $S_2(\pi\pi^*)$ state at time delay = 0 consists of two peaks, one at slightly lower binding energy than the ground state, due to N3 ionization (408.38 eV), and the other peak due to N1 ionization (410.24 eV). However, our experimental spectra do not show the latter peak, which may be due to insufficient instrumental resolution (see Figure~\ref{fig:FIG.3}(a), blue dashed line).\\ 
\\
\noindent
For longer time delays of 150 fs and 400 fs, theory predicts that the spectral signature of the $S_1(n\pi^*)$ state consists of a single, broadened peak at slightly lower binding energy than the ground state. This is indeed visible in Figure~\ref{fig:FIG.3}(b) as an intensity increase in range A, and depletion in range B.

\section{Discussion}
We begin by discussing the time constants for the decay of the $S_2(\pi\pi^*)$ and $S_1(n\pi^*)$ states, as determined from the experiment, and then relate them to the decay mechanisms revealed by the simulations. In experimental studies, $S_2$ and $S_1$ are often used to denote the diabatic $\pi\pi^*$ and $n\pi^*$ states, though this is not strictly correct: $S_1$ refers to the first excited state, and $S_2$ to the second excited state. We use these labels in combination with the diabatic notation $\pi\pi^*$ and $n\pi^*$ for consistency with earlier work. 

\begin{figure*}[!ht]
\includegraphics[width = 6 in]{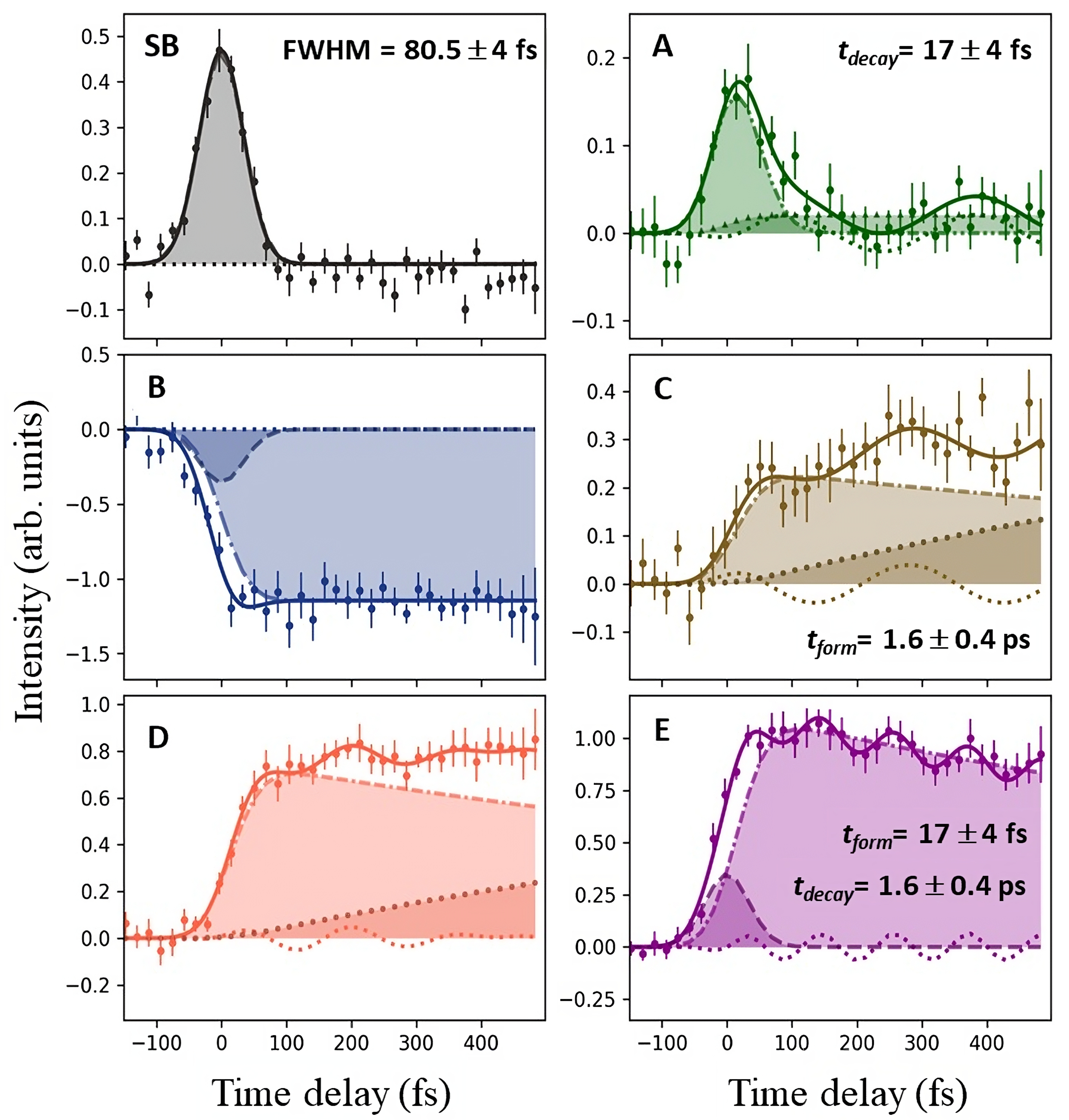}
\caption{O 1s photoelectron intensity integrated over BE regions SB, and A to E (see Figure~\ref{fig:FIG.2}(b)) as a function of time delay. Time decay/formation constants are extracted from the global fit 
(see section~S3, SI) 
%(see section~\ref{SI-section:3}, SI) 
of the O 1s data. Shaded areas represent the population dynamics of the corresponding states identified by the fit. The oscillations (square dots, line) in the O 1s signal intensities for the ranges A, C, D and E are also shown.} 
\label{fig:FIG.4} 
\end{figure*}

The photoelectron intensities integrated over the marked regions in Figure~\ref{fig:FIG.2}(b) were plotted as a function of time. A global fitting analysis was applied to extract the time constants, with the results shown in Figure~\ref{fig:FIG.4} 
%(see section~\ref{SI-section:3} for more details, SI). 
(see section~S3 for more details, SI). 

Briefly, the model assumes a Gaussian excitation function, followed by an exponential decay describing the conversion of $S_2(\pi\pi^*)$ to
$S_1(n\pi^*)$, where the decay time constant also corresponds to the formation time of the $S_1$ state. 
For the portion of the population that undergoes direct deactivation to the electronic ground state, the exponential decay of the diabatic $\pi\pi^*$ state represents the formation of the hot $S_0$ state. Theoretical simulations show that the time spent by trajectories in the $S_1(\pi\pi^*)$ state after passing through the $S_2/S_1$ CoIn (see dashed red arrow in Figure~\ref{fig:FIG.1}(c)) is very short, well below the experimental time resolution. 
Consequently, this time interval was excluded from the fitting procedure. 
An additional exponential decay describes the intersystem crossing of the $S_1(n\pi^*)$ state to the two lowest triplet $T_1/T_2$ states.

The lower binding energy sideband (see Figure~\ref{fig:FIG.4}, range SB) was fitted with a Gaussian profile and yielded a cross-correlation time of 80.5 $\pm$ 4 fs.

Within this model, the intensity of the $S_2$ state has an average decay time constant of
17 $\pm$ 4 fs (see Figure~\ref{fig:FIG.4}, range A), much shorter than in most previous measurements \cite{ullrich2004electronic,yu2016internal,canuel2005excited,kang2002intrinsic,matsika2013ultrafast}. Despite this time constant being smaller than the response function of the instrument, we are able to capture this with high precision due to the high signal-to-noise level of the measurement and the good spectral separation between the signals representing the $S_2$  and $S_1$ populations. Our value is model dependent, and, as such, its accuracy depends on the correctness of the model. Nevertheless, it agrees exceptionally well with the value of 17 $\pm$ 1 fs reported by Miura et al.\cite{miura2023formation}

The computed O 1s spectra at 0, 10, 25, and 50 fs show that as the molecule evolves on the $S_2(\pi\pi^*)$ surface, the signal at 536.0 eV gradually shifts towards higher binding energies. By 50 fs, the signal becomes indistinguishable from the ground state signal (see Figure~\ref{fig:FIG.2}(a)). Although this timescale is longer than the "best estimate" of 12.5 fs for the decay of $S_2$ obtained by Matsika and coworkers using XMS-CASPT2-based surface hopping simulations,\cite{chakraborty2021faraday} the underlying direct relaxation mechanism $S_2(\pi\pi^*) \rightarrow S_1(\pi\pi^*) \rightarrow S_0$ is the same. The direct pathway of $S_2(\pi\pi^*)$ decay 
%(see Figure~\ref{SI-fig:FIG.16}, 
%section~\ref{SI-section:9}, SI), 
(see Figure~S16, section~S9, SI),
leads to a population of vibrationally excited molecules in the electronic ground state {\it{via}} an ethylenic-type seam of CoIns, that involves a twist around the C5=C6 bond.\cite{lan2009photoinduced,fingerhut2014probing,chakraborty2021time,chakraborty2021faraday,milovanovic2021simulation,richter2014ultrafast}. 

To identify the spectral signature of the vibrationally excited molecules, we calculated the O 1s ground state spectra for the trajectories that returned back to the $S_0$ state, synchronizing them to start from the respective $S_1(\pi\pi^*)/S_0$ CoIns at $t=0$. Indeed, the results 
(see Figure~S13, SI) 
%(see Figure~\ref{SI-fig:FIG.13}, SI) 
show a shift in the O 1s ground state peak toward higher binding energies, consistent with the observed broadening of the main peak and the increased intensity between 538-539 eV (see 
Figure~\ref{fig:FIG.2}(b)), which we attribute to the hot ground state (see discussion below, and section~S3, SI).
%section~\ref{SI-section:3}, SI).
 
For the $S_1$ state, we extracted a single decay constant of 1.6 $\pm$ 0.4 ps (see Figure~\ref{fig:FIG.4}, range E). The trapping of the population in the $S_1(n\pi^*)$ state within the indicated time constant is in agreement with several previous experimental studies. \cite{kang2002intrinsic,canuel2005excited,yu2016internal,chakraborty2021time,miura2023formation} However, unlike Miura et al.\cite{miura2023formation},
we find no evidence of the $S_1(n\pi^*)\rightarrow S_0$ relaxation mechanism, as the ground state depletion signal does not recover in the measured O 1s spectra, even at very long times (see Figure~\ref{fig:FIG.2}(b)). Instead, we find the $S_1$ state to be the doorway to the triplet states (see Figure~\ref{fig:FIG.4}, range C-D)\cite{richter2014ultrafast,serrano2007intrinsic,climent2007intrinsic,etinski2009intersystem,karak2023photophysics}. 
To test this hypothesis, we computed the spectra for the two lowest triplet states, $^3n\pi^*$ and $^3\pi\pi^*$, at their respective minimum energy geometries. 
We note that the ordering of these states is geometry dependent and that each can stabilize as the lowest triplet state ($T_1$). 
Based on these computations (see Figure~\ref{fig:FIG.2}(a), orange and brown lines) the O 1s features observed in the BE range of 539-542 eV were assigned to the formation of the
$^3\pi\pi^*$ and $^3n\pi^*$ states (see Figure~\ref{fig:FIG.2}(b)). The observed ISC time constant of 1.6 $\pm$ 0.4 ps falls between the two constants of 0.44 ps and 3.48 ps reported by Miura et al. \cite{miura2023formation}. They suggested that the presence of these two constants may result from two dynamic processes, but could also arise from the dependence of the valence cross section on the electronic and structural properties of uracil.
Our calculations show that the core level spectrum varies little during the simulation time (see Figure~\ref{fig:FIG.2}(a) and 
Figure~S14 from SI), 
%Figure~\ref{SI-fig:FIG.14} from SI), 
due to motion on the $S_1(n\pi^*)$ potential energy surface. 
Therefore, it is more likely that the authors of Ref.~\citenum{miura2023formation} observed spectral changes resulting from variations in cross sections. 

In the second part of the discussion we examine the influence of the nuclear motion on the XPS signal. In the present O 1s spectra of uracil (Figure~\ref{fig:FIG.4}), clear oscillations have been observed, and we have performed a Fourier Transform analysis in order to extract their frequencies (Figure~S9, 
section~S3, SI). 
%(Figure~\ref{SI-fig:FIG.9}, 
%section~\ref{SI-section:3}, SI). 
In the time domain, Figure~\ref{fig:FIG.4}, oscillations in the O 1s signal intensities are depicted. One can see that the intensity oscillates in regions A ($S_2(\pi\pi^*$)), C ($S_1/T_1$), D ($S_1(n\pi^*$)) and E ($S_1(n\pi^*$)) resulting in different frequency components for each domain. In particular, the main frequency modes of 114.6 cm$^{-1}$ (291 fs), 114.8 cm$^{-1}$ (290.5 fs), 198.1 cm$^{-1}$ (168 fs) and 292.7 cm$^{-1}$ (114 fs) were identified for the A, C, D and E energy ranges, respectively 
(see Figure~S9, section~S3, SI). 
%(see Figure~\ref{SI-fig:FIG.9}, 
%section~\ref{SI-section:3}, SI).
We interpret these oscillations as being due to variations in the binding energy, rather than cross section, of the ionic states. At fixed kinetic energy, this results in an oscillation of the intensity.

\begin{figure*}[htb]
\includegraphics[width=6.5in]{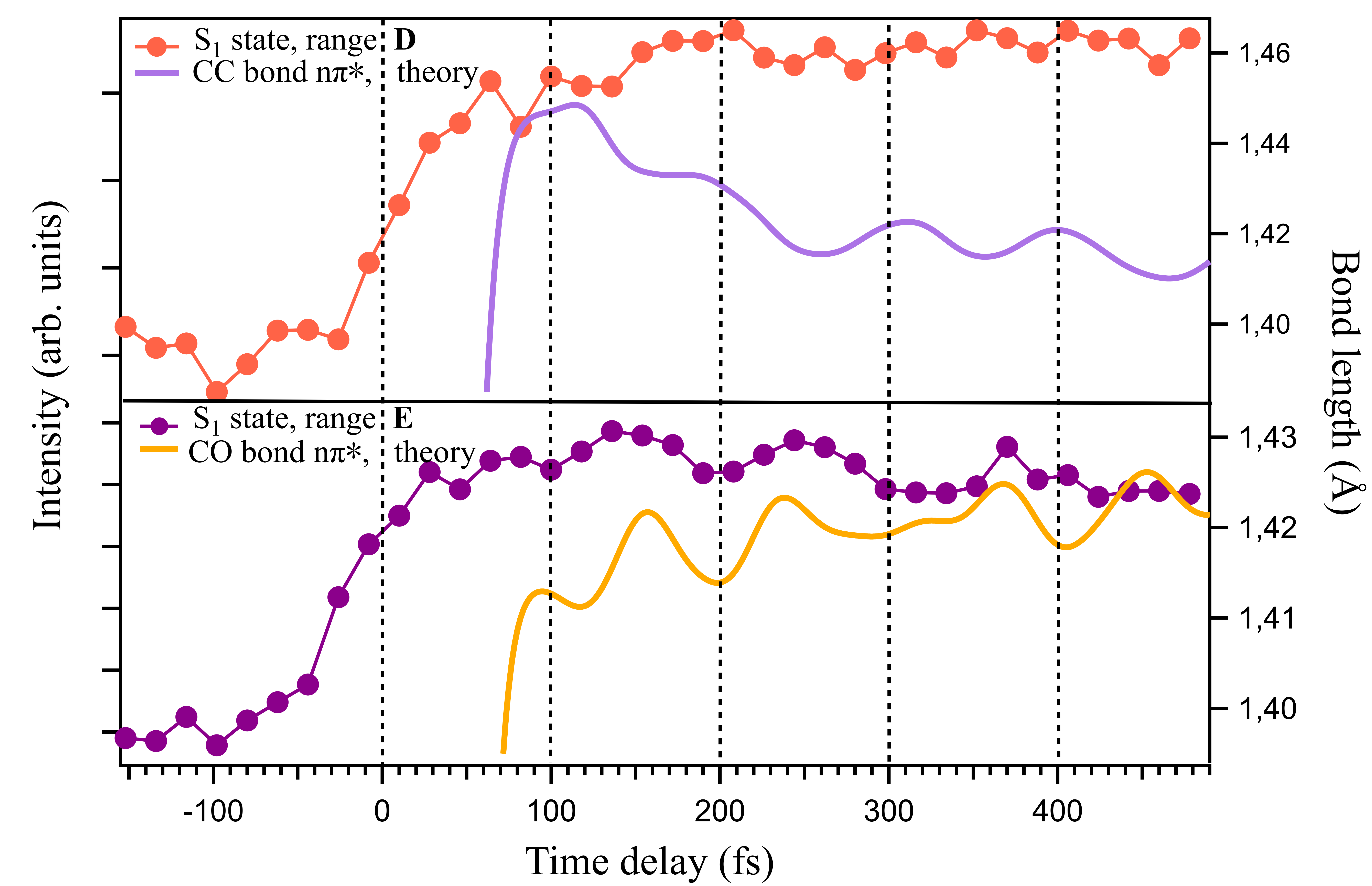}
\caption{Measured O 1s intensity in energy ranges D (orange dots)  and E (dark magenta dots) as a function of time delay, and calculated mean values of the C5=C6 (light purple line) and C4=O8 (yellow line) bond lengths for a set of trajectories in the $S_1(n\pi^*)$ state. Calculated curves were shifted by 30 fs in order to match experimental data.}  
\label{fig:FIG.5} 
\end{figure*}

The O 1s intensity in the energy ranges A and C oscillates with a similar frequency but in antiphase, and we assigned this effect to the HGS oscillations in this region, 
%(see section~\ref{SI-section:3}, SI), 
(see section~S3, SI), 
that is, the energies correspond to the wings of the hot ground state spectrum. As for the oscillations in the ranges E and D, we correlate them to the calculated bond lengths demonstrating very satisfactory agreement with theory (Figure~\ref{fig:FIG.5}). 
To be specific, we analyzed the normal modes and computed the average lengths of the C5=C6 and C4=O8 bonds (see section S10, SI). This approach is more accurate than comparing vibrational frequencies of the ground state, as the $\pi \to \pi^*$ excitation causes a sudden $\sim$ 0.1 \text{\AA} extension of the two double bonds  
%(see Figure~\ref{SI-fig:FIG.17}, 
%section~\ref{SI-section:10}, SI). 
(see Figure~S17, section~S10, SI). 
Since the stretching of these bonds does not correspond to specific normal modes, multiple modes involved in the stretching become activated.

At thermal energies, vibrational effects on core spectra are usually weak, but in the present case, the ground electronic state of the molecule is vibrationally excited (equivalent to a temperature of about 1800 K, assuming equipartition), and much of this energy is concentrated in the C5=C6 and C4=O8 bonds. The large-amplitude vibrations then modulate the core spectrum. As an example, in Figure~\ref{fig:FIG.5}, the O 1s intensity for the energy ranges D and E in Figure~\ref{fig:FIG.2}(b) is plotted as a function of time delay together with the calculated C5=C6 and C4=O8 mean bond lengths for the ensemble of trajectories that reached the $S_1(n\pi^*)$ state. Most of the maxima and minima in the intensity and bond lengths correlate, and the two pairs of curves are roughly anti-correlated. 
These results indicate that the fluctuations in intensity are directly related to vibrational motion: when the C4=O8 bond length increases, the intensity in region D also increases. 
Simultaneously, the intensity in region E decreases, and the C5=C6 bond length decreases. The extremely large nuclear motion  affects the electronic wave functions and this is reflected in core binding energies.

\begin{figure*}[htb]
\includegraphics[width = 6.5 in]{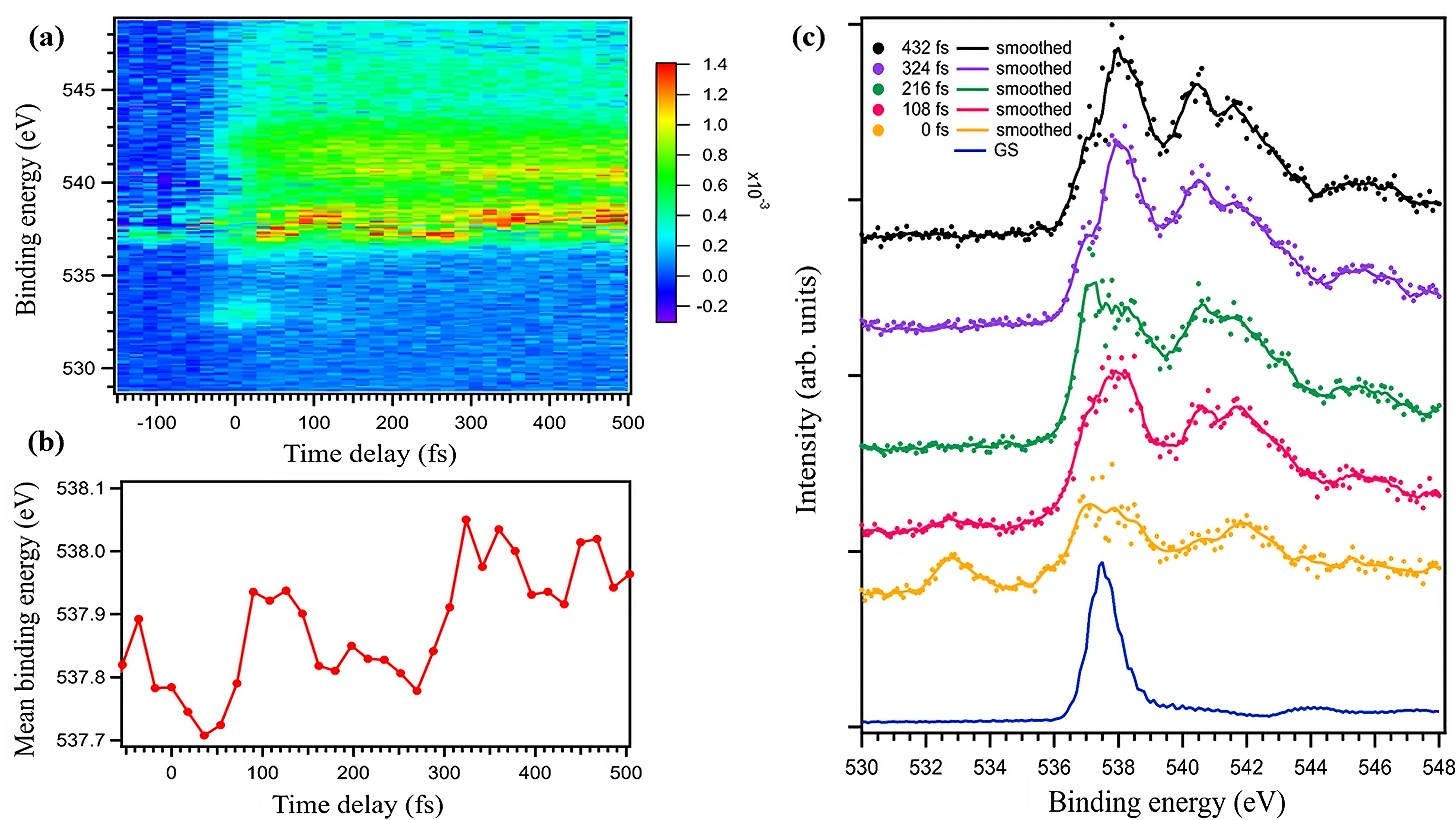}
\caption{ (a) False color map of O 1s intensity after scaled subtraction, as a function of binding energy and time delay. (b) Mean value of the main line binding energy with scaled subtraction. (c) Bottom curve: spectrum of the ground state (GS). Upper curves: spectra of the excited state, centered at the indicated time delays and integrated over time intervals of 108 fs. The spectra have been smoothed by a 5-point box function, corresponding to 0.5 eV resolution.} 
\label{fig:FIG.6} 
\end{figure*}

Miura et al. \cite{miura2023formation} also observed an oscillatory signal attributed to photoionization from the $n\pi^*$ state to a presumably excited valence state of the uracil ion. 
Their Fourier analysis indicated vibrational frequencies of 135 and 315 \textrm{cm}$^{-1}$ i.e. periods of 247 and 95 fs, respectively. These periods are quite close to the values we observed, but the authors\cite{miura2023formation} could not relate them to specific normal modes. In a very recent paper, Karashina and Suzuki \cite{karashima2025exploring} associated the frequency 300 \textrm{cm}$^{-1}$ (110 fs period) with the motion of the C5-H moiety, whereas we assign a similar feature to C5=C6 and C4=O8 bond stretching. The same authors reported vibrational coherence transfer during the ultrafast internal conversion from the $S_2(\pi\pi^*$) to the $S_1(n\pi^*$) state of gaseous uracil. The prominent peak at 750 \textrm{cm}$^{-1}$ (44.5 fs) was found in the frequency spectra and assigned to the breathing mode of the aromatic ring, characterized by a large change in the N$_1$C$_6$C$_5$ angle.\cite{borrego2021tracking} This frequency is not accessible in our data due to our lower temporal resolution.

Furthermore, as noted above, we performed scaled subtraction of the O 1s spectra to generate spectra of the $S_1$ state without the negative features due to depletion of the ground state. 
The method is described in the Supporting Information 
%(see section~\ref{SI-section:2}), 
(see section~S2),
and we found a value of $f=0.068$, corresponding to 6.8\% of the initial ground state population transferred to the $S_1$ state. 
The approximations in this approach are rather crude, as the presence of the hot ground state is not fully accounted for, but the method serves to highlight some features. 
For example, Figure~\ref{fig:FIG.6}(a) shows a map generated with the procedure, and there are clearly time varying features in the energy region of the main line (core ionized ionic state). Its mean energy is shown in 
Figure~\ref{fig:FIG.6}(b) as a function of time delay, and it oscillates with a period of about 200 fs. In addition, the TR-XPS spectra generated by integrating over time intervals of 108 fs demonstrate distinct changes of shape as a function of time (see Figure~\ref{fig:FIG.6}(c)).

The main line shows a shoulder at lower binding energy which becomes strongest at 216 fs and then weakens, while the average energy shows an underlying trend of increasing binding energy. 
The oscillations in binding energy may be due to O7 and O8 in the hot ground state, or to O7 in the $S_1$ state, since both contain substantial internal energy which is in the process of being thermalised. At higher binding energies, the structures unequivocally due to $S_1$ also show changes. 
In the spectrum at time zero, the peak at 542 eV is initially dominant, and during relaxation, the peak at 540.6 eV becomes dominant. These data confirm the prediction of our calculations that the core spectra are not constant in time, but oscillate during the relaxation process.
%%%%%%%%%%%%%%%%%%%%%%%%%%%%%%%%%%%%%%%%%%%%%%%%%%%%%%%%%%
%%            CONCLUSIONS 
%%
%%%%%%%%%%%%%%%%%%%%%%%%%%%%%%%%%%%%%%%%%%%%%%%%%%%%%%%%%%
\section{Conclusions}
In this study, time-resolved X-ray photoelectron spectroscopy (TR-XPS) was employed to directly probe the oxygen and nitrogen atoms in uracil, 
 uncovering its relaxation dynamics following UV excitation. 
While all core levels contributed to understanding the dynamics, the O 1s signal emerged as particularly sensitive. Our findings demonstrate that TR-XPS spectra capture both charge migration and structure changes in the excited states. The present experimental data are in good agreement with computed XPS spectra for the ground and excited states of uracil performed on structures derived from surface hopping non-adiabatic dynamics simulations.

The photoinduced dynamics of uracil take place on three-time scales: (i) an ultrafast step, (ii) an intermediate step of around 2 ps, and (iii) a slow relaxation, greater than 10 ps. 
The first interval, along with the ultrafast formation of vibrationally excited ground state molecules, provides evidence of the direct $S_2(\pi\pi^*) \to S_1(\pi\pi^*) \to S_0$ deactivation pathway. The indirect deactivation channel involves two steps: a fast $S_2(\pi\pi^*) \to S_1(n\pi^*)$ prompt internal conversion occurring with a time constant 17 $\pm$ 4 fs and the second step $S_1(n\pi^*) \to T_1(\pi\pi^*)$ lasting for a time constant of 1.6 $\pm$ 0.4 ps. This confirms that the $n\pi^*$ state is the doorway for ISC relaxation to the triplet states, responsible for the slower ps/ns dynamics of uracil.

The results provided by the present work clearly demonstrate that TR-XPS spectra are also capable to follow nuclear dynamics  effects. The observed oscillations in the O 1s signal intensity (at fixed kinetic energy) are directly related to the vibrational motion of the C5=C6 and C4=O8 bond lengths during the ultrafast electronic relaxation processes of uracil.

\section{Acknowledgement}

We acknowledge the European XFEL in Schenefeld, Germany, for the provision of x-ray free-electron laser beam time at the SQS instrument and thank the EuXFEL staff for their assistance. MD, DF and CV acknowledge financial support under the National Recovery and Resilience Plan (NRRP), Mission 4, Component 2, Investment 1.1, Call for tender No.and 104 published on 2.2.2022 by the Italian Ministry of University and Research (MUR), funded by the European Union – NextGenerationEU–  Project 2022K5W59T DEMIST – CUP B53D23013780006 - Grant Assignment Decree No.~958 adopted on 30/06/2023 by the Italian Ministry of Ministry of University and Research (MUR) and Call for tender No. 1409 published on 14.9.2022 by the Italian Ministry of University and Research (MUR), funded by the European Union – NextGenerationEU– Project P20224AWLB "HAPPY" – CUP B53D23025210001 - Grant Assignment Decree No.~1386 adopted on 01/09/2023 by the Italian Ministry of University and Research (MUR). SC acknowledges support from the Novo Nordisk Foundation Data Science Research Infrastructure 2022 Grant: A high-performance computing infrastructure for data-driven research
on sustainable energy materials, Grant No.~NNF22OC0078009,
and from Hamburg's Cluster of Excellence "CUI: Advanced Imaging of Matter", 
2024 Mildred Dresselhaus Prize.
Part of the work was carried out with support from the European Union’s Horizon 2020
Research and Innovation Programme under the Marie Sk{\l}odowska-Curie Individual Fellowship (BNCT, Grant Agreement No.~101027796). MS and ND acknowledge the support by the Croatian Science Foundation under the project
numbers [HRZZ-IP-2020-02-9932 and HRZZ-IP-2022-10-4658].

\section{Supporting Information}

Experimental and theoretical carbon K-edge TR-XPS spectra of uracil 
%(see section~\ref{SI-section:12}). 
(see section~S12).
Details concerning experimental data processing
and time-resolved fitting. Additional figures, tables, and structures that support conclusions presented in the article.

%%%%%%%%%%%%%%%%%%%%%%%%%%%%%%%%%%%%%%%%%%%%%%%%%%%%%%%%%%%%%%%%
%%%%%%%%%%%%%%%%%%%%%%%%%%%%%%%%%%%%%%%%%%%%%%%%%%%%%%%%%%%%%%%%%%%%%
\section{Data availability}

The raw data recorded for the experiment at the European XFEL are available at: 
10.22003/XFEL.EU-DATA-003014-00. 

%The data that support the findings of this study will be openly available following an embargo at the following URL/DOI: (metadata DOI). Data will be available from DD MM YYYY.

%%%%%%%%%%%%%%%%%%%%%%%%%%%%%%%%%%%%%%%%%%%%%%%%%%%%%%%%%%%%%%%%
\bibliography{References}
\end{document}

\maketitle
\clearpage
\tableofcontents
\clearpage
\section{~Supporting figures for the experimental section.}
\label{section:1}
\begin{figure}[htb]
\includegraphics[width = 5.5 in]{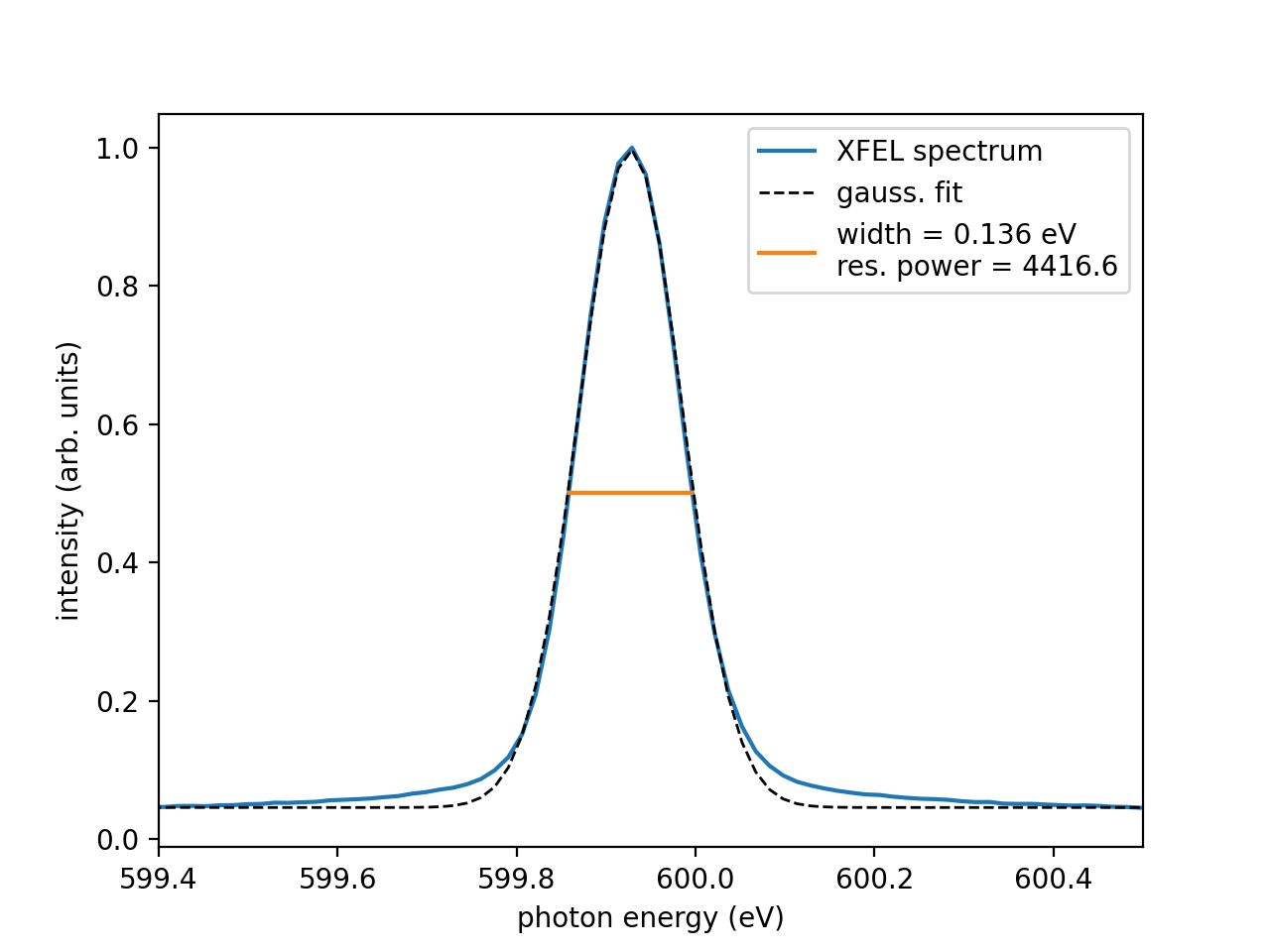}
\caption{FEL spectrum as the output of the SQS instrument monochromator measured at the spectrometer operation mode at 600 eV. Average of 236 shots. A fit with a Gaussian function is shown.}
\label{fig:FIG.1} 
\end{figure}

The experiment was conducted at the Small Quantum Systems (SQS) instrument located at the SASE3 undulator of the European XFEL \cite{mazza2023beam}. 
The XFEL beam consisted of 10 trains of pulses per second, with 166 pulses per train at a repetition rate of 376 kHz within the train. 
The undulator was tuned to provide X-ray pulses centered at a photon energy of 600 eV with a 5 eV full width at half maximum (FWHM) bandwidth and mean pulse energy of 6.8 mJ.  The SQS instrument monochromator \cite{gerasimova2022soft} was used to reduce the FEL bandwidth to \text{0.136 eV }(FWHM) with a measured resolving power of E/$\Delta$E=4.4$\times$\text{10}$^{3}$ (see Fig.~\ref{fig:FIG.1}). 

\begin{figure}[htb]
\includegraphics[width = 5.5 in]{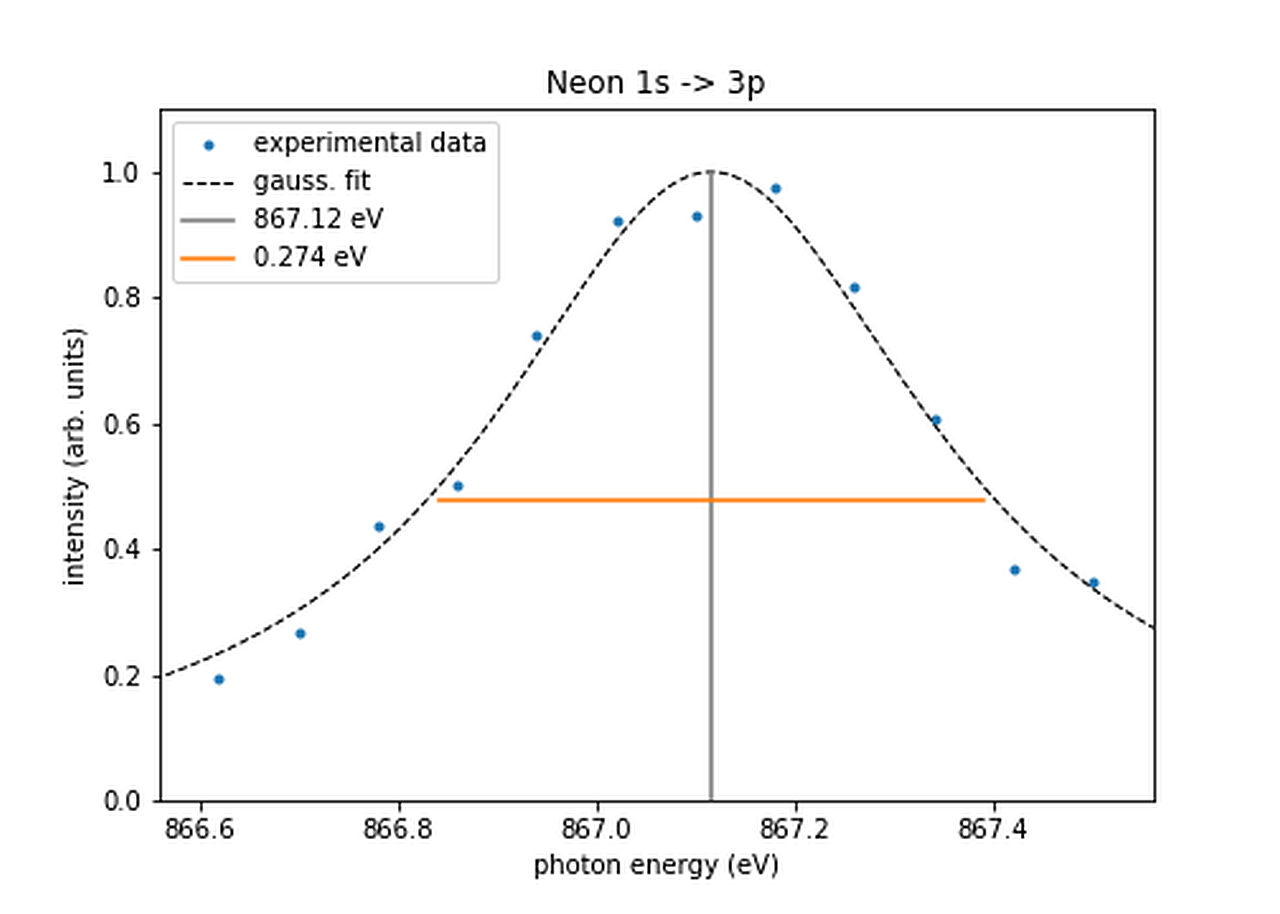}
\caption{Ne 1s–3p X-ray absorption spectrum used to calibrate the monochromator. An offset of 0.44 eV with respect to the nominal energy was found.}
\label{fig:FIG.2} 
\end{figure}

The same monochromator is also utilized as a spectrometer for spectral diagnostics of the FEL beam. The spectrometer operation mode is realized by introducing a YAG:Ce crystal into the focal plane of the monochromator \cite{koch2019operation}. 
Figure~\ref{fig:FIG.2} shows the well-known atomic Ne 1s–3p X-ray absorption resonance at 867.12 eV, used to  calibrate the monochromator energy scale\cite{coreno1999measurement}.

\begin{figure}[htb]
\includegraphics[width = 6.5 in]{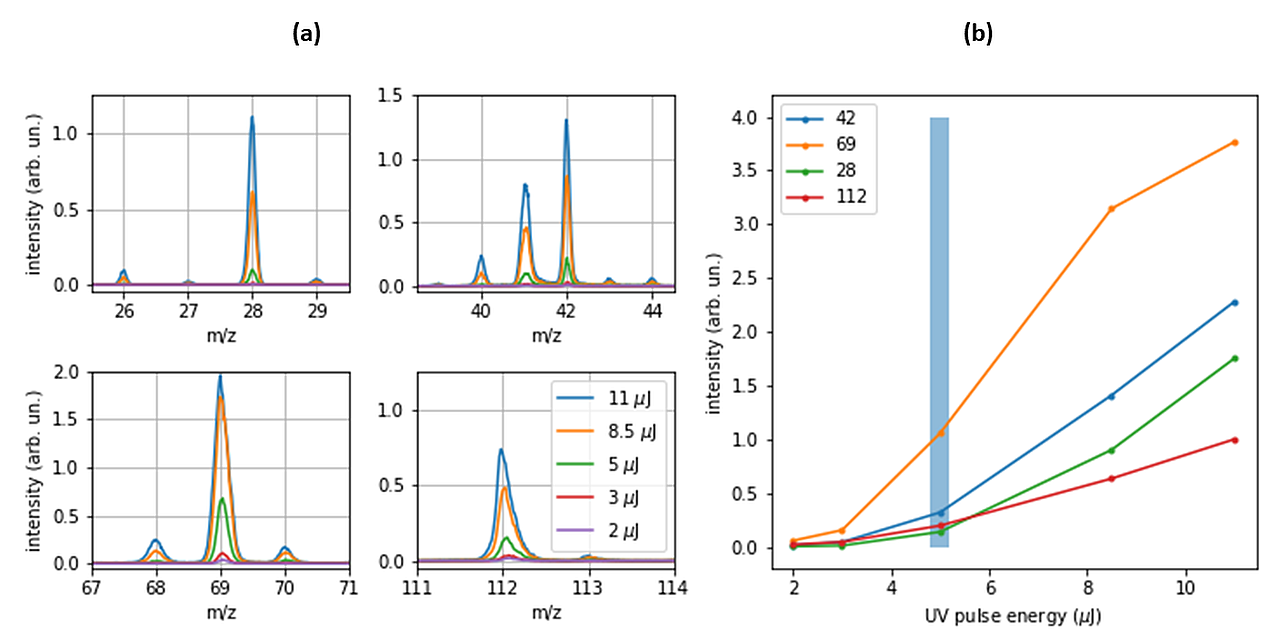}
\caption{(a) Measured TOF spectra of uracil as a function of UV pulse energies (only main fragments are shown). 
(b) UV dependence of signal for the selected fragments. The blue bar indicates the operating optical laser energy.}
\label{fig:FIG.3} 
\end{figure}

Uracil was excited to its lowest-energy $\pi\pi^*$ absorption band \cite{milovanovic2021simulation}
by ultrashort 264 nm pulses (75 fs) focused to a diameter of 150 $\mu$m (FWHM). During the experiment, photoion time-of-flight (TOF) mass spectra were measured with different laser intensities in order to check that the molecules were not excessively pumped by the UV laser. 
The parent ion showed a quadratic dependence on the laser intensity, as expected for weak pumps. 
Some of the fragments (e.g., mass 69) showed saturation at UV pulse energy > 8$\mu$J. Hence, in the present experiment the UV pump energy was set to 5$\mu$J in order to avoid the fragment saturation region (see Fig.~\ref{fig:FIG.3}).

\begin{figure}[htb]
\includegraphics[width = 6 in]{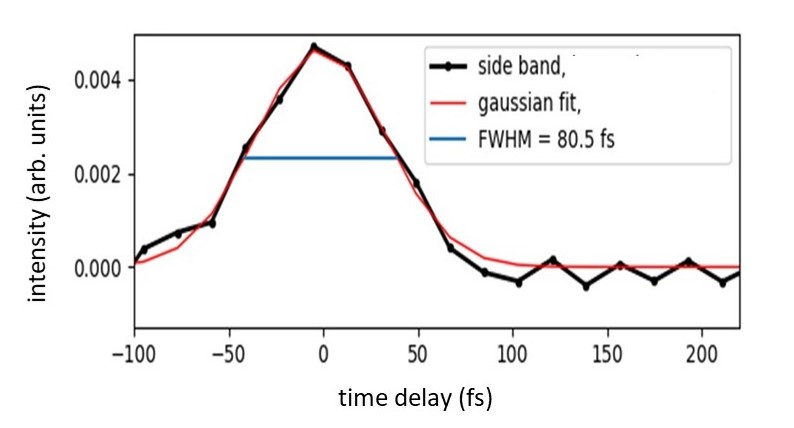}
\caption{The cross-correlation function between the UV optical and FEL laser pulses.}
\label{fig:FIG.4} 
\end{figure}

The temporal and spatial overlap of the XFEL and UV lasers were optimized using the depletion of the Auger signal of uracil as target signal. 
The depletion was induced by photo-fragmentation at full UV energy (\text{11} $\mu$J per pulse). 
During the pump-probe measurements, the presence of sideband (SB) signal occurring in TR-XPS spectra around t=0 delay was always ascertained in order to ensure the temporal overlap between both UV and X-ray pulses. 
In addition, the SB signal yielded the precise pump–probe instrument response function resulting in a value of 80.5 fs for the cross-correlation of the temporal widths of the pump and probe 
(see Fig.~\ref{fig:FIG.4})).

Uracil was purchased from Sigma Aldrich and evaporated without further processing from an effusive capillary oven \cite{mcfarland2014experimental} at a temperature of 160 $^\circ$C  into an ultra-high vacuum chamber, creating a molecular beam that interacts with two photon beams (optical and X-ray) in the center of a magnetic bottle electron spectrometer (MBES). This spectrometer provides highly efficient time-of-flight measurements of electrons emitted into 4$\pi$ solid angle, with nearly 100 $\%$ angular acceptance. Its operating principle is similar to other units described elsewhere. \cite{kruit1983magnetic,hikosaka2014high,borne2024design}

The MBES was calibrated by scanning the photon energy so that the Ne 1s photoline \cite{coreno1999measurement} fell in the same kinetic energy ranges as those used for the O 1s, N 1s and C 1s photolines at 600~eV photon energy. 
Three different retardation voltages of $-45$ V, $-180$ V and $-302$ V were applied to the electrostatic lens of the MBES in order to obtain high resolution photoelectron spectra at the oxygen, nitrogen and carbon K-edges, respectively.

\begin{figure}[htb]
\includegraphics[width = 6 in]{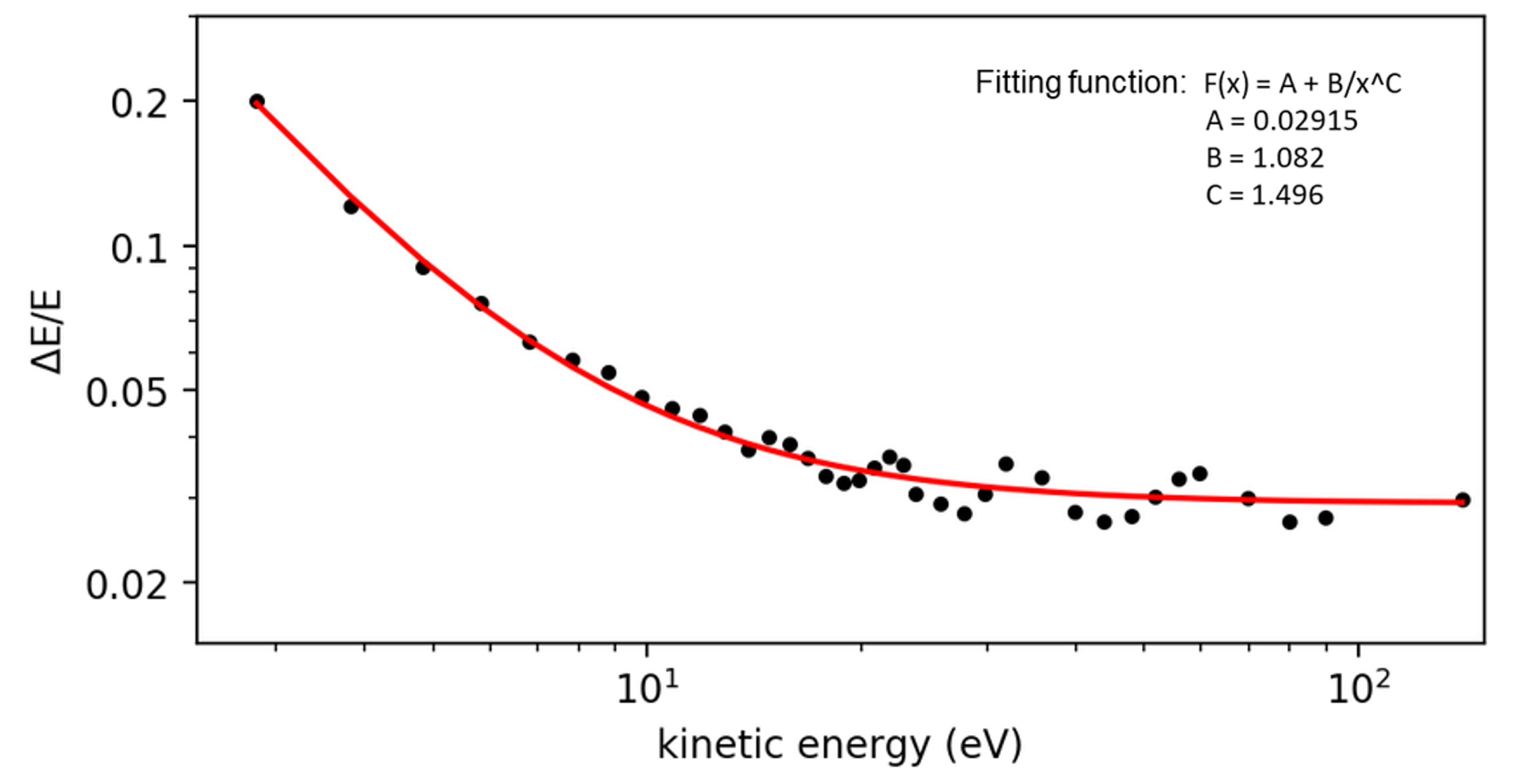}
\caption{Measured relative resolution ($\Delta E$/$E$) as a function of kinetic energy $E$ after the retardation.}
\label{fig:FIG.5} 
\end{figure}

The resolution of the MBES was estimated from the photoionization of Ne 1s (binding energy 870.2 eV), by scanning the photon energy of the XFEL in the energy region between 911 eV and 1048 eV. 
A retardation of 45 V was applied. Figure~\ref{fig:FIG.5} shows the measured relative resolution $\Delta E$/$E$ ($\Delta E$ is the FWHM and $E$ is the kinetic energy as measured, i.e., not compensated for the retardation potential) as a function of $E$ which was fitted with the equation:
\begin{equation}
F(x)= A+B/x^C
\label{eq:fit}
\end{equation}
where \textit{x} is the kinetic energy and \textit{F} is the relative resolution ($\Delta E /E$). 
The asymptotic relative resolution for kinetic energies $>$ 20 eV was estimated to be $\Delta E/E \approx$ 0.03.

\cleardoublepage
\section{~Procedure for scaled subtraction.}
\label{section:2}
We analysed the O 1s spectra using scaled subtraction of the ground state spectrum, to generate the spectra of the $S_1$ state without the features due to depletion of the ground state. 
The latter appear as negative structures when the ground state spectrum is subtracted from the excited state spectra without taking account of depletion. 
The raw data also contained an O 1s peak due to residual water, which subtracts out when the difference is taken without scaling. 
For scaled subtraction, a residue of this peak may remain in the spectrum as a negative feature, so the data were corrected by fitting the water peak and subtracting it. 
We know from calculation that the O8 1s binding energy increases strongly in the $S_1$ state, while the O7 1s binding energy changes very little.

We integrated the intensity of the ground state spectrum over the main line (O7 and O8) in the range 536 to 539.6 eV and over the shake-up range, from 539.6 to 549 eV, and found the ratio of intensities was 0.71:0.29. 
We then assumed that in the $S_1$ state, half of the intensity in the ground state main line was transferred to the shake-up region due to the binding energy shift of O8. 
Assuming the shake-up intensity is unchanged in the $S_1$ state, a ratio of 0.355:0.645 is expected for the main line (O7) to shake-up plus O8 intensities. 
The ground state spectrum was then scaled and subtracted from the excited state spectra; the scaling factor was chosen to give the expected ratio, and had a value of $f=0.068$. 
As an example, Figure 6(a) 
(see main text) 
shows a map generated with this procedure, where time-varying features in the energy region of the main line are clearly observed. 
In particular, it is evident that the mean energy of the main line oscillates as a function of time with a period of about 200 fs, Figure 6(b), main text.

\cleardoublepage
\section{~Procedure used for fitting the O 1s TR-XPS spectra of uracil.}
\label{section:3}
We modeled the observed dynamics by a set of ordinary differential equations that describe the evolution of populations in each state over time. We indicate with $g(t, t_0, \sigma)$ the effective temporal envelope of the UV laser pulse triggering the observed dynamics, where $t_0$ is the pump-probe time zero and $\sigma$ the standard deviation of the pulse. 
This term incorporates the cross-correlation of the pump and probe pulses, capturing both the excitation profile and the experimental temporal resolution. 
It is assumed to be a Gaussian:

\begin{equation}
g(t, t_0, \sigma) = \frac{\exp\left(-\frac{(t - t_0)^2}{2 \sigma^2}\right)}{\sigma \sqrt{2 \pi}}
\end{equation}

The population dynamics of the ground state $S_0(t)$,
%, the hot ground state $HGS(Tùt)$
the initially excited state $S_2(t)$, the lowest-energy excited state $S_1(t)$, and the triplet state $T(t)$ are described by the following set of differential equations:
\begin{eqnarray}
%\begin{equation}
%\begin{cases}
\frac{dS_0(t)}{dt} = -g(t, t_0, \sigma) \nonumber\\
   % \frac{HGS(t)}{dt} &= \frac{S_2(t)}{t_{3}} \\ 
\frac{dS_2(t)}{dt} = g(t, t_0, \sigma) - \frac{S_2(t)}{t_{1}}  \nonumber\\
\frac{dS_1(t)}{dt} = \frac{S_2(t)}{t_{1}} - \frac{S_1(t)}{t_{2}} \nonumber\\
\frac{dT(t)}{dt} = \frac{S_1(t)}{t_{2}}
%\end{cases}
\label{eq:model_diff}
%\end{equation}
\end{eqnarray}
where $t_{1}$ is the decay time from the singlet $S_2$ to $S_1$ state, and $t_2$ is the decay time from the $S_1$ to the triplet $T$ state. Eq.~\eqref{eq:model_diff} was solved numerically using the \texttt{solve\_ivp} function from the \texttt{scipy.integrate} library. The free parameters $t_1$, $t_2$, $\sigma$, and $t_0$ were optimized through a global fitting procedure to reproduce the time-dependent dynamics observed in the energy ranges presented in Figure 2(b) up to 500 fs (main text).

\begin{figure}[htb]
\centering 
\includegraphics[width=0.9\textwidth]{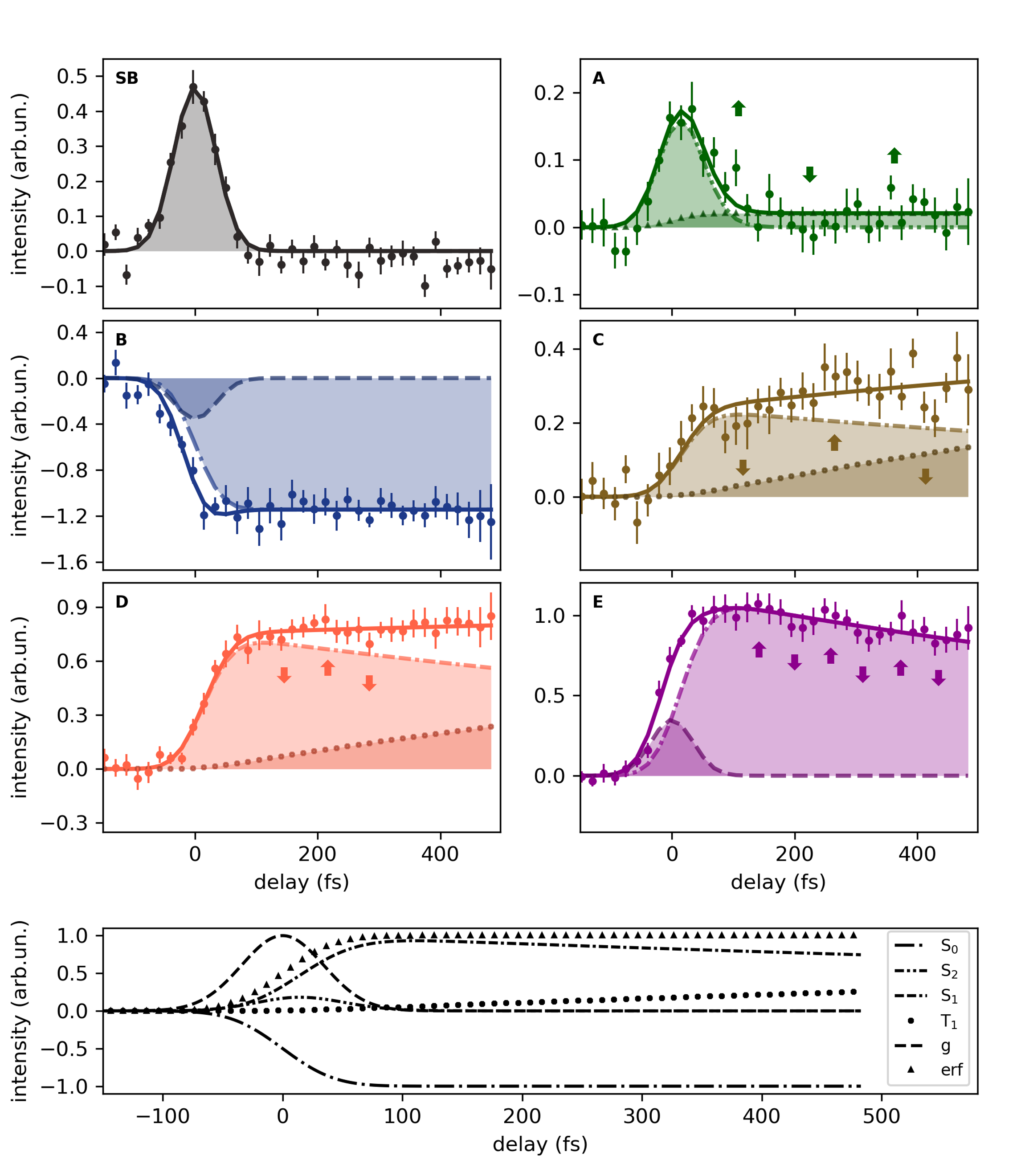}
\caption{Experimental pump-probe differential yield for the six energy ranges SB (sideband), A, B, C, D, and E. The data points and standard errors are shown as dots and error bars, respectively. 
The solid lines represent the results of the fit to the data. For energy ranges A, B, C, D, and E, where two contributions are considered (see Eq.~\ref{eq:model}), both contributions, scaled by the fitted amplitude, are shown. 
The line styles of these contributions are defined in the legend of the bottom panel. 
Additionally, the best-fit solution to the differential equation for each state is plotted in the bottom panel, following the same line style coding. 
Up and down arrows in panels A, C, D and E indicate the observed clear deviations of the experimental data from the fit.}
\label{fig:FIG.6}
\end{figure}

Figure~\ref{fig:FIG.6} shows the result of the fit for the six energy regions, while Table~\ref{tab:table1} summarizes best-fit values of all parameters. For the global fit, $N_j = 6$ time-dependent functions $f_j(t)$ were considered, with $j = \mathrm{SB} 
~(\mathrm{sideband}), \mathrm{A}, \mathrm{B}, \mathrm{C}, \mathrm{D}, \mathrm{E}$ indicating the energy ranges over which the signal was integrated (see Figure 2(b) of the main manuscript). 
Specifically, $f_\mathrm{SB}(t)$ was obtained by integrating in the range $531.9-534$ eV, $f_\mathrm{A}(t)$ in the range $535.2-536.6$ eV, $f_\mathrm{B}(t)$ in the range $537.2-538$ eV, $f_\mathrm{C}(t)$ in the range $539-540$ eV, $f_\mathrm{D}(t)$ in the range $540.1-541.5$ eV, and $f_\mathrm{E}(t)$ in the range $541.6-543.5$ eV. 

\begin{table}[ht]
\centering
\begin{tabular}{|c|c|c|}
\hline
\textbf{Parameter} & \textbf{Fitted Value} & \textbf{Error} \\
\hline
$t_0$ (fs) & 81.14 & 1.7 \\
$\sigma$ (fs) & 34.26 & 1.4 \\
$t_1$ (fs) & 17.03 & 3.7 \\
$t_2$ (fs) & 1585.40 & 377.5 \\
$a_{SB, g}$ & 0.47 & 0.022 \\
$a_{A, S_2}$ & 0.88 & 0.18 \\
$a_{A, erf}$ & 0.021 & 0.005 \\
$a_{B, S_0}$ & 1.15 & 0.023 \\
$a_{B, g}$ & 0.36 & 0.058 \\
$a_{C, S_1}$ & 0.24 & 0.023 \\
$a_{C, T}$ & 0.53 & 0.14 \\
$a_{D, S_1}$ & 0.76 & 0.034 \\
$a_{D, T}$ & 0.93 & 0.20 \\
$a_{E,S_1}$ & 1.12 & 0.045 \\
$a_{E, g}$ & 0.35 & 0.045 \\
\hline
\end{tabular}
\caption{Fitted parameters and their errors. The amplitudes and amplitude errors are shown in arbitrary units, but they have all been scaled by the same common factor to show the relative weight that each state has in a certain energy region.}
\label{tab:table1}
\end{table}

Guided by the calculated spectra shown in Figure 2(a) (main text), only specific states were considered for each band, and the contribution (amplitude) of each state was fitted accordingly. The amplitude contribution of the $i$-th state to the $j$-th energy range is denoted as $a_{j, i}$. The following equations describe the time-dependent contributions for each energy range:
\begin{eqnarray}
f_\mathrm{SB}(t)  & =& a_{\mathrm{SB}, g} \cdot g(t) \nonumber\\
f_\mathrm{A}(t)  & =& a_{\mathrm{A}, S_2} \cdot S_2(t) + a_{\mathrm{A}, \mathrm{erf}} \cdot \mathrm{erf}(t) \nonumber\\
f_\mathrm{B}(t)  & =& - a_{\mathrm{B}, g} g(t) + a_{\mathrm{B}, S_0} S_0(t) \nonumber\\
f_\mathrm{C}(t)  &=& a_{\mathrm{C}, S_1} S_1(t) + a_{\mathrm{C}, T} T(t) \nonumber\\
f_\mathrm{D}(t)  &=& a_{\mathrm{D}, S_1} S_1(t) + a_{\mathrm{D}, T} T(t) \nonumber\\
f_\mathrm{E}(t)  &= & a_{\mathrm{E}, S_1} S_1(t) + a_{\mathrm{E}, g} g(t) 
\nonumber\\
\label{eq:model}
\end{eqnarray}
where the following considerations were made. 
The sideband at zero delay is expected to produce a negative cross-correlation signal for band $\mathrm{B}$ and a positive cross-correlation signal for bands $\mathrm{SB}$ and $\mathrm{E}$. 
The triplet state $T$ primarily contributes to bands $\mathrm{C}$ and $\mathrm{D}$, where it becomes dominant at delays exceeding 500 fs. The singlet state $S_1$ is expected to contribute to bands $\mathrm{C}$, $\mathrm{D}$, and $\mathrm{E}$, while $S_2$ mainly affects band $\mathrm{A}$. An additional contribution for band $\mathrm{A}$ is considered, proportional through the constant $a_{\mathrm{A}, \mathrm{erf}}$ to the error function:
\begin{eqnarray}
\mathrm{erf}(t, t_0, \sigma) &=& \int_{-\infty}^t g(t, t_0, \sigma) dt \label{eq:erf}
\end{eqnarray}
The inclusion of this term is justified by the fact that the signal associated with band $\mathrm{A}$ does not average to zero after the $S_2$ decays, which, as discussed later, can be associated with the presence of a hot ground state (HGS) long-lived population signature in this region. Note that the contribution of HGS was not directly taken into account by the model of Eq.~\eqref{eq:model_diff}.

The fitting procedure was carried out using the \texttt{minimize} function from the \texttt{scipy.optimize} library, employing the Nelder-Mead optimization algorithm. The chi-squared error function to minimize was defined as:

\begin{equation}
\chi^2= \sum_j^{N_j} \sum_t^{N_t} \left(\frac{\bar{f}_j(t) - f_j(t; \{p_k\})}{ \sigma_{j, t} }\right)^2
\end{equation}
where the sum over $t$ runs and over the $N_t = 38$ temporal bins, $\sigma_{j,t}$ represents the standard error for band $j$ at time $t$ (shown as error bars in Fig.~\ref{fig:FIG.6}), $\bar{f}_j(t)$ is the measured yield for band $j$ at time $t$ (shown as dots in Fig.~\ref{fig:FIG.6}), and $f_j(t)$ is the function modeling band $j$ at time $t$. 
The function $f_j(t)$ is parametrized by the set of $N_k = 15$ parameters $p_k$ listed in Table \ref{tab:table1}, which  include $t_0$, $\sigma$, $t_1$, $t_2$, and the set of amplitudes $\{a_{j,i}\}$ specific to the $j$-th band, as defined in Eqs.~\eqref{eq:model}.
The starting parameters were initialized to physically reasonable values and constrained to be positive. The optimization process terminated successfully 
%with a chi-squared valueat the end of the fit being $197.11$. The
with the reduced chi-squared value $\chi_r^2 = \chi^2  /({N_t\cdot N_j-N_k}) = 0.9$, which, being very close to one, indicates a very good agreement between the model and the experimental data. 
The errors in the retrieved parameters were estimated by calculating the numerical Hessian matrix of the error function.  
The inverse of the Hessian matrix of the error function provides an estimate of the parameter covariance matrix, from which the standard deviations of the parameters are extracted. 
The formula for the error of each parameter, $\Delta p_k$, is:

\begin{equation}
\Delta p_k = \sqrt{\left( \text{Cov}(p_k, p_k) \right)} = \sqrt{\left( \left( 2 \mathbf{H}^{-1} \right)_{kk} \right)}~,
\end{equation}
where $\mathbf{H}^{-1}$ is the inverse of the Hessian matrix, and $\left( \mathbf{H}^{-1} \right)_{kk}$ represents its diagonal elements, which are equal to half the variances of the model parameters. 
These errors reflect the uncertainty associated with each of the fit parameters.

\begin{figure}[htb]
\centering 
\includegraphics[width=0.9\textwidth]{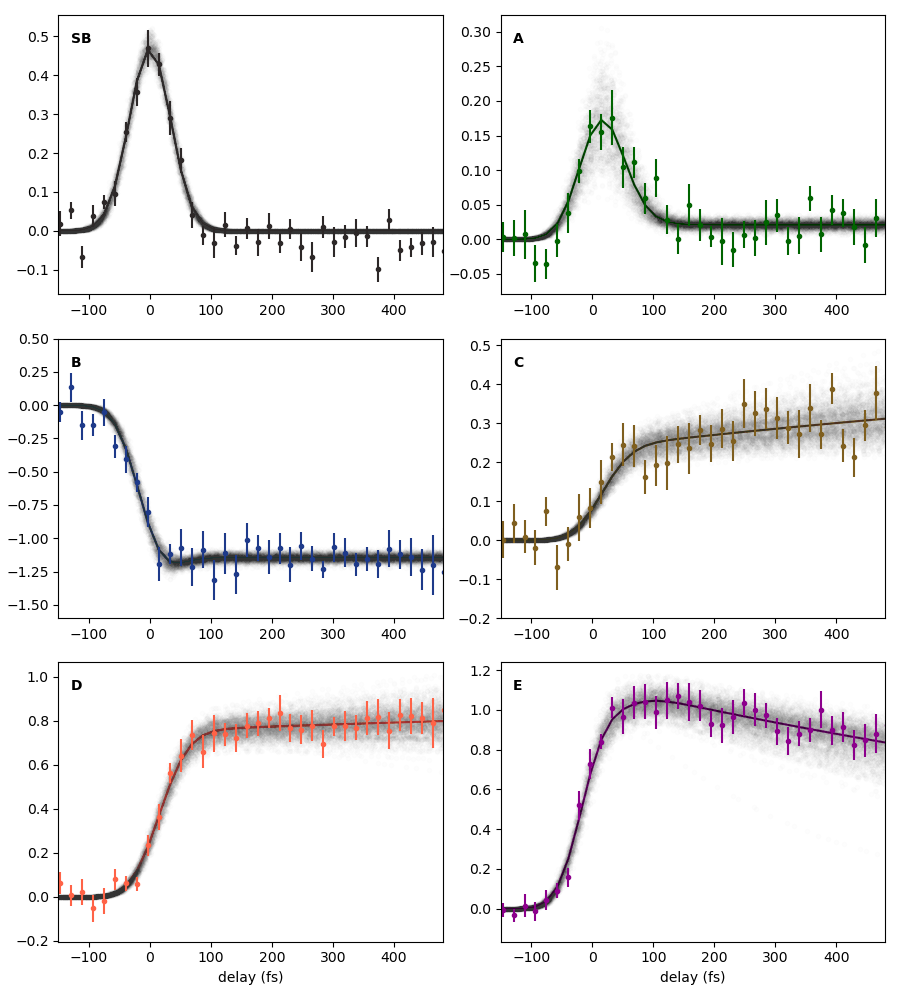}
\caption{Experimental pump-probe differential yield for the six energy ranges $\mathrm{SB}$ (sideband), $\mathrm{A}$, $\mathrm{B}$, $\mathrm{C}$, $\mathrm{D}$, and $\mathrm{E}$. The data points and standard errors are shown as dots and error bars, respectively. 
The solid lines represent the results of the fit to the data. The gray shaded scatter dots represent the values obtained from the 500 parameter samples, each drawn from a Gaussian distribution with means and standard deviations corresponding to the values listed in Table~\ref{tab:table1}, and propagating these sample through the model.}
\label{fig:FIG.7}
\end{figure}

To better understand the impact of these errors on the time-dependent contributions $f_j(t)$, we propagated the retrieved uncertainties through the model. This was achieved numerically by generating 500 parameter samples, each drawn from a Gaussian distribution with means and standard deviations corresponding to the values listed in Table~\ref{tab:table1}. The resulting uncertainty plots for each $f_j(t)$ are shown in Fig.~\ref{fig:FIG.7} as gray shaded scatter plots, and provide a visual representation of the confidence intervals associated with the fitted contributions.

As it is possible to observe from Fig.~\ref{fig:FIG.7}, the uncertainty associated with the model generally overlaps with the experimental uncertainties, except in a few specific regions that will be addressed later. This visual agreement is consistent with the reduced chi-squared value being close to one, indicating a good fit between the model and the experimental data and further confirming the validity of the retrieved mean values and standard errors reported in Table~\ref{tab:table1}.

One remarkable result is the very low value of $t_1$ obtained from the fit, which is 17~$\pm~$4$\,$fs, despite the cross-correlation between the pump and probe pulses being approximately 81 fs. At first glance, this result may seem surprising.
The key to interpret this result lies in the fact that the model incorporates the sideband signal, which is proportional to the cross-correlation function and represents the instrument response function (IRF) of the system. Since the IRF is known, the experimental data can, in principle, be deconvolved from the IRF. This allows for a resolution that surpasses the FWHM of the IRF. Such an approach is
well documented in the literature, where fitting techniques or deconvolution techniques combined with a knowledge of the IRS have been shown to achieve significantly higher temporal resolution than the nominal instrumental limit. \cite{iagatti2015photophysical,fazel2023fluorescence,mouton2010deconvolution}

In our specific case, the high accuracy achieved for the $t_1$ parameter arises from two key factors. First, we have a precise knowledge of the IRS, enabled by the direct observation of the sideband signal and the low signal-to-noise ratio in this region. Second, we can directly observe the rise time of the $S_2$ state from band $\mathrm{D}$, which is not accessible in conventional visible pump-probe spectroscopy. As we will show soon, these two pieces of information alone are sufficient to determine the value of $t_1$, which means that the decay observed in band $\mathrm{A}$, which is noisier and overlaps with the HGS, has a smaller influence on the retrieval of $t_1$. The model further refines this result by enforcing the condition that the decay time of band $A$ and the rise time of band $\mathrm{D}$ are the same.

\begin{figure}[htb]
\centering 
\includegraphics[width=0.9\textwidth]{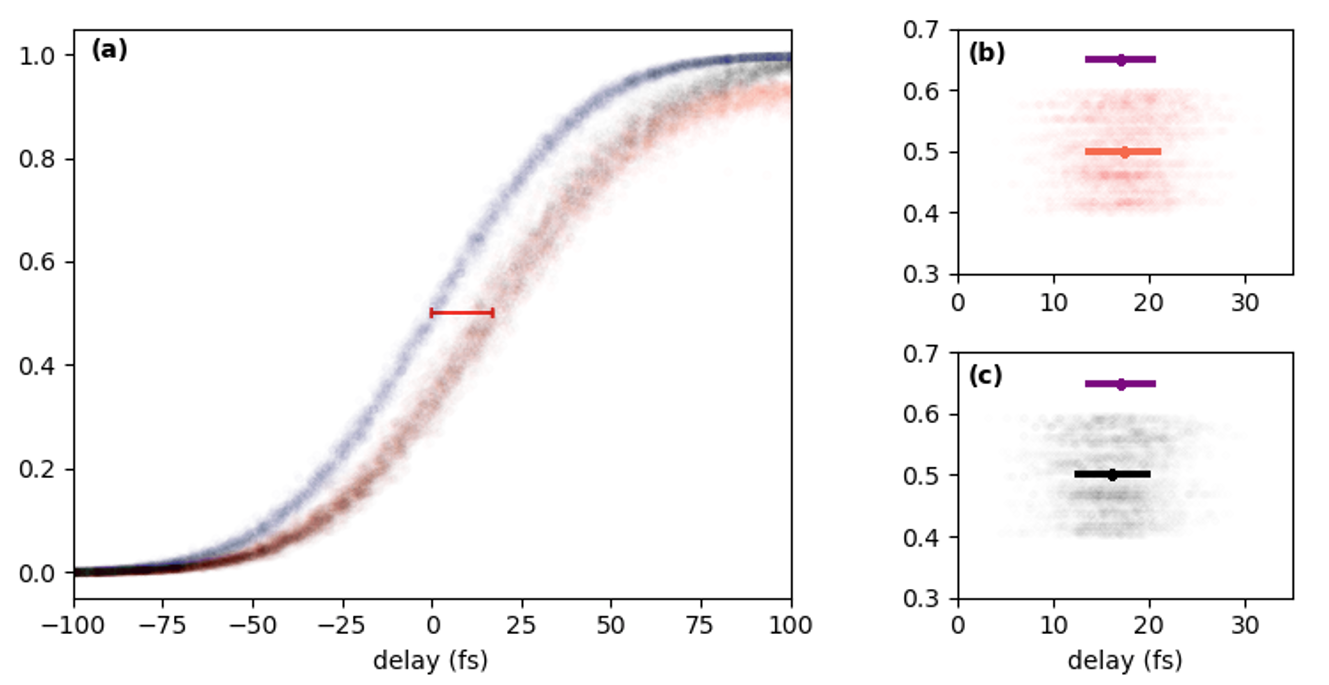}
\caption{(a) Comparison of $\mathrm{erf}(t)$ (blue shaded dots) and $\hat{f}_D(t) = f_D(t)/a_{D, S_1}$ (orange shaded dots), obtained from 500 samples generated using the mean and standard errors listed in Table~\ref{tab:table1}. The shift between the rise times of these two functions at half-maximum provides an estimate of $t_1$. The horizontal red bar corresponds to 17 fs. Also shown is $\hat{S}_1(t)$ (black shaded dots) for the same 500 samples, which is the rising contribution of $S_1$ ($t_2 \rightarrow \infty$). 
(b) Delays in the amplitude range of 0.4 to 0.6 of $\hat{f}_D(t)$ with respect to $\mathrm{erf}(t)$ (red shaded dots). The mean and standard deviation is represented by the red error bar. (c) Delays in the amplitude range of 0.4 to 0.6 of $\hat{S_1}_D(t)$ with respect to $\mathrm{erf}(t)$ (black shaded dots). The mean and standard deviation is represented by the black error bar. In (b) and (c), the purple error bar indicates the value obtained from the model for comparison.}
\label{fig:FIG.8}
\end{figure}

To visually illustrate this concept, Fig.~\ref{fig:FIG.8}(a) shows a comparison between $\mathrm{erf}(t)$ (blue shaded dots) and $\hat{f}_D(t) = f_D(t)/a_{D, S_1}$ (orange shaded dots), both obtained from 500 samples generated using the mean and standard errors listed in Table~\ref{tab:table1}. In the regime where $t_2 \gg t_1$, which applies to our case, the shift between the rise times of these two functions at half-maximum provides already a good estimate of $t_1$. As shown in the figure, this shift is clearly distinguishable and emerges above the model error.
To quantify this shift, we computed the cross-difference (sample by sample) of the delays within the amplitude range of 0.4 to 0.6, as indicated by the red shaded dots in Fig.~\ref{fig:FIG.8}(b). The mean and standard deviation of these differences, represented by the red error bar, yields a value of $17.28 \pm 4.18\,\mathrm{fs}$. This result is consistent with the value obtained from the model, shown by the purple error bar in the same subplot.
To exclude the contribution of the decay to the triplet state, we applied the same procedure by directly comparing $\mathrm{erf}(t)$ with the solution of the differential equation ${d\hat{S}_1(t)}/{dt} = S_2 / t_2$, which represents the rising contribution of $S_1$ or, equivalently, the solution for $S_1$ when $t_2 \rightarrow \infty$. The $S_1$ obtained from the same 500 parameter samples is shown by black shaded dots in Fig.~\ref{fig:FIG.8}(a), which only start do deviate from $\hat{f}_D(t)$ in the region of $t > 60\,$fs. The difference of the delays in the amplitude range of 0.4 to 0.6 is indicated by the black shaded dots in Fig.~\ref{fig:FIG.8}(c). The mean and standard deviation of these differences, represented by the black error bar, yield a value of $16.15 \pm 4\,\mathrm{fs}$. This result is slightly lower than the one obtained from the model (shown in purple in Fig.~\ref{fig:FIG.8}(c))  but still within the error margin. Nevertheless, the higher $t_1$ obtained from the model could be ascribed to an effect of band $\mathrm{A}$, which slightly increases its value compared to the one obtained when only band $\mathrm{D}$ is considered.

Despite the goodness of the fit, some clear disagreements were observed, especially for the curves in panels A, C, D and E. 
These deviations are highlighted by the up and down arrows in Fig.~\ref{fig:FIG.6}, which indicate the direction of the offset between the experimental data and the fit result. The deviations show a periodic behavior and may have two different origins. 

The energy regions A and C are close to the main ground state O8 and O7 peaks. 
The population of a HGS has been revealed by the scaled subtraction procedure (see main text and section S2). 
The population of an HGS with binding energy oscillating around band B affects the signal around neighboring energy bands A and C. 
This is seen as an anti-correlated signal, A at lower kinetic energy and C at higher kinetic energy with respect to B (see arrow directions in Figs.~\ref{fig:FIG.6} A and C). In principle, the population transfer to the HGS should be included in the model of Eq.~\ref{eq:model}. However, a clear formation time for the HGS could not be identified, while its presence, as shown later, is revealed by the Fourier analysis of these oscillations. 
For bands D and E, fast oscillations were attributed to the modulation of a normal mode in the excited $S_1$ state, and the observed deviations with respect to the fit are indicated by the arrows in the corresponding panels (see Fig.~\ref{fig:FIG.6}). 

\begin{figure}[htb] 
\centering \includegraphics[width=0.9\textwidth]{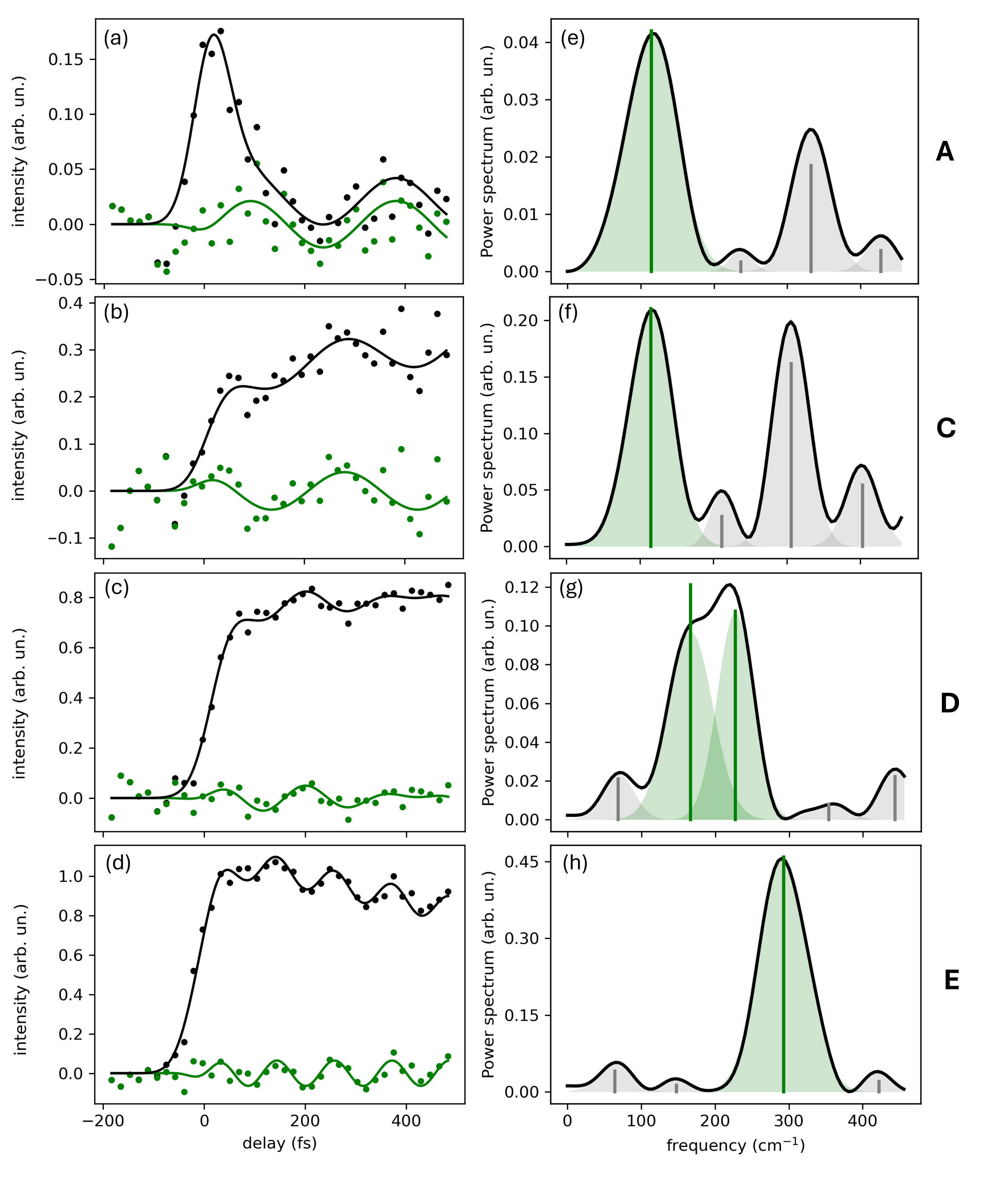}  
\caption{Panels (a-d) show the experimental data (black dots) for the energy bands corresponding to Fig.~\ref{fig:FIG.6} panels A, C, D, and E. The green dots represent residuals obtained by subtracting the fit from the experimental data. The oscillations retrieved via Fourier analysis are shown as green lines, while the black line represents the sum of the fit and the retrieved oscillation. Panels (e-h) display the power spectrum of the residuals (black), with shaded areas representing the Gaussian modes identified by the multi-Gaussian fit. Each mode’s central frequency is marked by a vertical line proportional to its power. Modes contributing to the green oscillations in panels (a-d) are highlighted in green.} 
\label{fig:FIG.9} 
\end{figure}

Taking these considerations into account, we estimated the period, phase, and amplitudes of the main oscillations observed in ranges $\mathrm{A}$, $\mathrm{C}$, $\mathrm{D}$, and $\mathrm{E}$ of Fig.~\ref{fig:FIG.6} using Fourier analysis. 
The experimental data for these bands are shown again as black dots in Figs.~\ref{fig:FIG.9}(a-d), corresponding to panels $a$, $b$, $c$, and $d$, respectively. 
The residuals, obtained by subtracting the previously obtained fit from the experimental data, are displayed as green dots in the same panels.

To isolate the region of positive delays, the residuals were multiplied by the error function (see Eq.~\eqref{eq:erf}). 
Subsequently, a Fourier Transform (FT) was performed, and the resulting power spectrum is presented in Figs.~\ref{fig:FIG.9}(e-h). The spectrum was fitted with a series of Gaussian modes, represented by shaded areas, whose central frequencies are marked by vertical lines with heights proportional to the mode power. 
The most relevant modes are highlighted in green: the oscillations in the time domain associated with these modes, after multiplication by the error function (Eq.~\eqref{eq:erf}), are represented by the green line in Figs.~\ref{fig:FIG.9}(a-d).

Starting from band $\mathrm{C}$, we detect two main frequency components at 114.8 cm$^{-1}$ (290.5 fs) and at 305.7 cm$^{-1}$ (109.1 fs) (see Fig.~\ref{fig:FIG.9}(f)). The low frequency mode produces the oscillatory behaviour represented by a green line in Fig.~\ref{fig:FIG.9}(b), and is responsible for the disagreement observed in Fig.~\ref{fig:FIG.6} (panel C). 
Also for band $\mathrm{A}$ we observe two main modes, very close in frequency to the ones detected for band $\mathrm{C}$, the lower one being at 114.6 cm$^{-1}$ (291.1 fs) and the higher at 332.3 cm$^{-1}$ (100.4 fs). 
The low frequency mode, the most intense, produces the oscillatory behavior represented by the green solid line in Fig.~\ref{fig:FIG.9}(a). The two oscillatory signals shown in Figs.~\ref{fig:FIG.9}(a) and (b) are approximately out of phase ($\Delta \Phi \approx 1.3\pi$), as anticipated, confirming that they can be associated with the HGS binding energy oscillations around band $\mathrm{B}$, which influence the signals in adjacent bands $\mathrm{A}$ and $\mathrm{C}$.

For band $\mathrm{D}$, two intense modes at 166.3 cm$^{-1}$ (200.6 fs) and 227 cm$^{-1}$  (146.9 fs) produce a beating with a central frequency of 198.1 cm$^{-1}$ (168.4 fs), shown as a green line in Fig.~\ref{fig:FIG.9}(c). For band E, a single dominant mode at 292.7 cm$^{-1}$ (113.9 fs) generates the oscillatory behavior depicted as a green solid line in Fig.~\ref{fig:FIG.9}(d). 
The shorter period of oscillation for band $\mathrm{E}$ compared to $\mathrm{D}$ is in good agreement with our theoretical calculations (see section S10). Additionally, at around 200 fs the oscillations of $\mathrm{D}$ and $\mathrm{E}$ are in antiphase, as predicted by the theory.

\cleardoublepage
\section{~Calculation of the time-resolved O 1s and N 1s spectra of uracil.}
\label{section:4}
Time-resolved  O 1s and N 1s XPS spectra of uracil were calculated using an ensemble of 48 SH (surface-hopping) trajectories. \cite{milovanovic2021simulation} Table \ref{tab:trajector} provides an overview of the distribution of SH trajectories between the $\pi\pi^*$ and $n\pi^*$ states at specific propagation times, illustrating the dynamic evolution of the ensemble.

\begin{table*}[!ht]
\caption{Distribution of an ensemble of 48 SH trajectories in the two valence excited states of uracil ($\pi\pi^*$ and $n\pi^*$) and the electronic ground state $S_0$ at specific delay times. 
The ensemble was used to compute the O 1s and N 1s spectra shown in Figure 2(a) and Figure 3(a), respectively (main text). 
The subset of trajectories that deactivate to $S_0$ contributes to the hot
ground state signal (see below, Fig.~\ref{fig:FIG.13}).}

\centering
\begin{tabular}{l|ccc}
\hline\hline
Time (fs) & $\pi\pi^*$  &  $n\pi^*$  & $S_0$ \\
\hline \hline\
0  &  48 &  0 & 0\\
10  &  48 & 0 & 0\\  
25  &  43 &  5 & 0 \\
50  &  40 &  7 & 1\\
150 &  15 &  23 & 10\\
450 &  0 &  31 & 17\\
\hline \hline
\end{tabular}
\label{tab:trajector}
\end{table*}

\begin{figure}[ht!]
\includegraphics[width = 5.5 in]{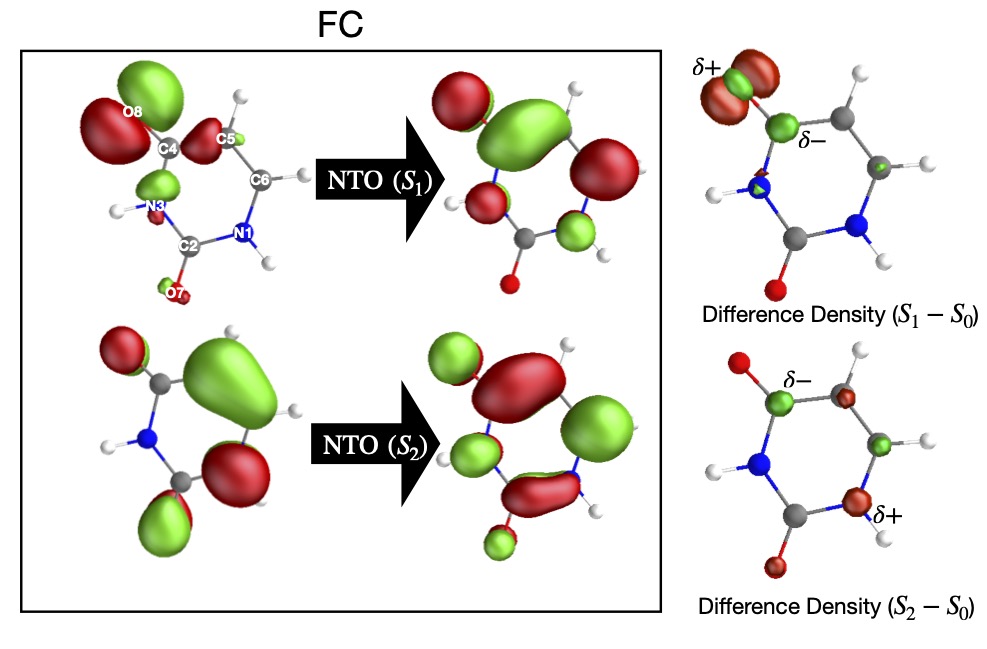}
\caption{Natural transition orbitals (NTOs) of the valence excited states $S_1 (n\pi^*)$ and $ S_2 (\pi\pi^*)$ computed at the ground state equilibrium geometry. The signs of partial charges "$\delta+/-$" are indicated, and numerical values are given in Table \ref{tab:charges}.}
\label{fig:FIG.NTO} 
\end{figure}

The partial atomic charges in the $S_2$ and $S_1$ excited states of uracil computed with RASPT2 from Mulliken population analysis are presented in Table \ref{tab:charges}. 
The transition to the $S_2(\pi\pi^*)$ state, which is conveniently described in terms of natural transition orbitals (NTOs) shown in Fig.~\ref{fig:FIG.NTO}, is accompanied by a flow of valence electrons. 
Specifically, electrons migrate from the bonding $\pi$ orbital primarily localized on the O8, O7, N1, and C5=C6 groups towards the antibonding $\pi^*$ orbital of the carbonyl group. 
Consequently, this electron redistribution leads to a migration of electron density from O7 and N1 towards the region encompassing N3 and the carbonyl group C4=O8. 
In contrast, the $S_2 \to S_1$ internal conversion leads to a significant reduction of the electron density on the O8 atom and a slight increase of electron density on the nitrogen atoms. 
On these grounds, the excess or deficit of valence electron density on oxygen or nitrogen is anticipated to result in a red-shift or blue-shift, respectively, of the corresponding core binding energies in the XPS spectra \cite{mayer2022following}.

\begin{table*}[!ht]
\caption{Partial charges from RASPT2 calculations at the FC geometry using the aug-cc-pVDZ basis set.
RASPT2 partial charges are given by the difference between the Mulliken charge of the corresponding atom in the excited state minus the Mulliken charge of the atom in the ground state (GS).}
\centering
\begin{tabular}{l|rrrrrrrr}
\hline\hline
 & O7 & O8  & N3 & N1 & C6 & C2 & C4 & C5  \\
\hline \hline
$\delta_{S_0}$ & $-$0.72 & $-$0.81 & $-$0.48 & $-$0.53 & 0.77 & 1.08 & 0.72 & 0.80 \\
$\delta_{S_1}$ & $-$0.66 & $-$0.47 & $-$0.54 & $-$0.55 & 0.68 & 1.10 & 0.63 & 0.72 \\
$\delta_{S_2}$ & $-$0.66 & $-$0.76 & $-$0.53 & $-$0.46 & 0.73 & 1.06 & 0.67 & 0.79 \\
\hline \hline
$\delta_{S_1 - S_0}$ &  0.06 & 0.34 & $-$0.06 & $-$0.02 & $-$0.09 & 0.02 & $-$0.09 & $-$0.08 \\
$\delta_{S_2 - S_0}$ &  0.06 & 0.05 & $-$0.05 &  0.07 & $-$0.04 & $-$0.02 & $-$0.05 & $-$0.01 \\
\hline \hline 
\end{tabular}\\
\label{tab:charges}
\end{table*}

Indeed, the experimentally observed intensity around 536 eV (see Figure 2(b), range A, main text) assigned to the core ionization of O8 in the $S_2$ bright state of uracil was found to be red-shifted with respect to the GS peak, due to the effect of an increase of electron density on the carbonyl group C4=O8 (see Figure 1(b) and Table \ref{tab:charges}). 
Conversely, the broad feature caused by the deficit of electron density on O8 in the $S_1(n\pi^*)$ state displays a notable blue-shift, as illustrated in the 2D map (see Figure 2(b), range C-E, main text). 
These experimental observations are in good agreement with our computations. 

As for the N 1s spectra, the computed partial atomic charges at the FC geometry show a significant decrease of the electron density at the N1 atom in the $S_2(\pi\pi^*)$ state and an increase at the N3 atom (see Table \ref{tab:charges}). 
Hence, a blue-shift relative to the GS is expected for the optically bright state in the N1s TR-XPS spectra. 
However, the signature of the bright state was only predicted theoretically, but not observed experimentally due to the lack of instrumental resolution (see Figure 3(a), blue dashed line, main text). 
In accord with our calculations, the partially positive charge on the N1 atom in the $S_2(\pi\pi^*)$ excited state in the FC region disappears for distorted geometries at 50 fs (gray), that is, for geometries at which the $S_2(\pi\pi^*)/S_1(n\pi^*)$ internal conversion takes place (see  Figure 3(a), main text). 
The simulated spectra for time delays of 150 and 400 fs, where signal is exclusively due to the dark state, demonstrate an additional shift towards lower binding energy compared to the ground state spectra. 
Indeed, the experimentally observed signature of the $S_1$ state of uracil was found to be red-shifted with respect to the GS, which is in line with the increase of electronic density on the N3 atom for this state (see Figure 3(b), range A, main text).

\begin{figure}[ht]
\includegraphics[width = 5.5 in]{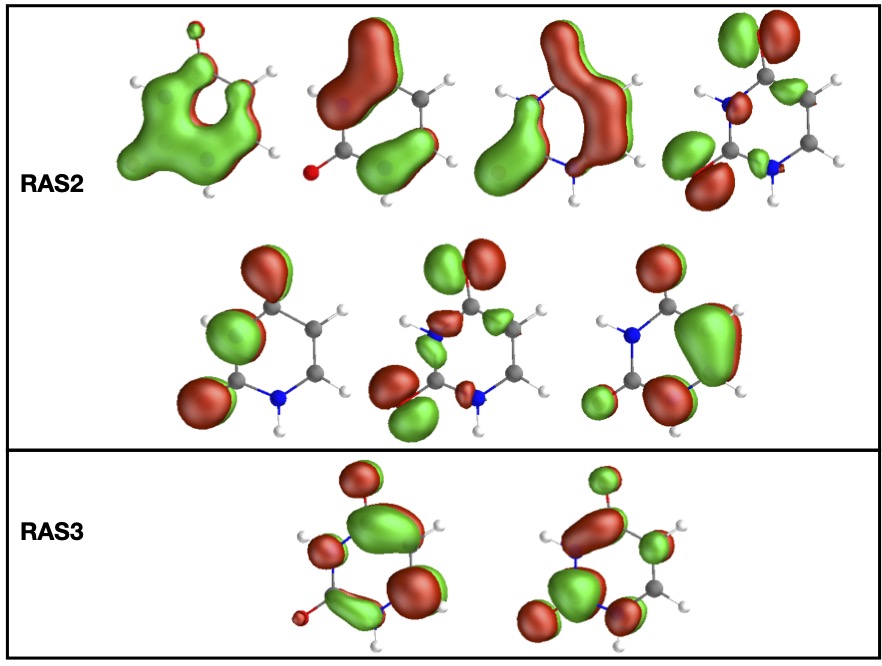}
\caption{Active space used in the RASSCF calculations in a three-fold division of the active space into RAS1, RAS2, and RAS3. The RAS1 subspace (not shown) consists of the pertinent core orbitals for which a single-hole configuration is enforced. The RAS2 subspace contains seven valence-occupied orbitals. 
Finally, the RAS3 subspace is formed by two $\pi^*$ orbitals and accepts a maximum of two electrons. 
The orbitals shown are state-averaged over ten neutral states.} 
\label{fig:FIG.11} 
\end{figure}

In addition, Fig.~\ref{fig:FIG.11} shows the active subspaces utilized in the restricted active space self-consistent field (RASSCF) calculations (see main text). RAS1 comprises the relevant core orbitals, RAS2 includes seven valence-occupied orbitals, and RAS3 is formed by two $\pi^*$ orbitals, each capable of accommodating a maximum of two electrons.

\cleardoublepage
\section{~Calculated O 1s spectra at the ground state equilibrium geometry.}
\label{section:5}
\begin{figure}[htb]
\centering
\includegraphics[width=4 in]{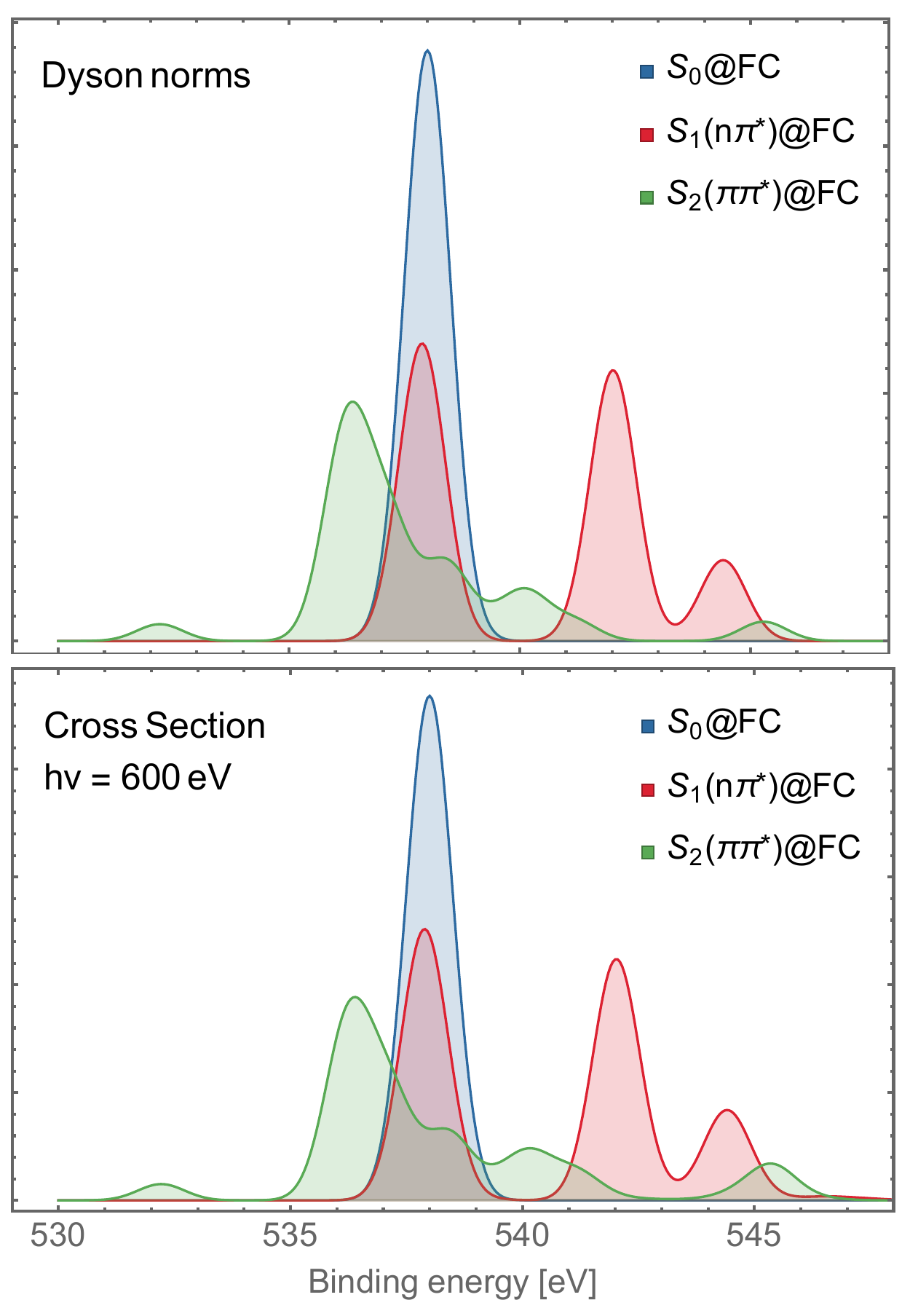}
\caption{Calculated O 1s XPS spectra using the equilibrium geometry of the ground state.Top panel: convolution of the computed ionization energies and Dyson norms with a Lorentzian function (FWHM = 0.4 eV). Bottom panel: spectrum obtained by convolution of the computed ionization energies and the cross sections computed for a photon energy of 600 eV. }
\label{FIG.12}
\end{figure}

O 1s XPS spectra calculated at the equilibrium geometry are shown in Fig.~\ref{FIG.12}. 
The cross sections have been obtained with an explicit description of the electronic continuum with an LCAO B-spline basis using the Tiresia code.\cite{TOFFOLI2024109038} 
Electronic exchange and correlation effects in the continuum have been accounted for by the LB94\cite{LB94} functional. 
The continuum states were described by a large one-center expansion of B-splines enclosed in a sphere of 25 a.u. with origin at the center of mass, using spherical harmonics of angular momentum up to $l_{max}=20$ to get well converged results. 
A small off-center expansion located over the nuclei varied from 0.5 to 1.0 a.u., larger for the heavier nuclei, and an angular expansion limited to $l_{max}=2$.

\cleardoublepage
\section{~Spectrum of the hot ground state (HGS).}
\label{section:6}
\begin{figure}[htb]
\includegraphics[width = 4.5 in]{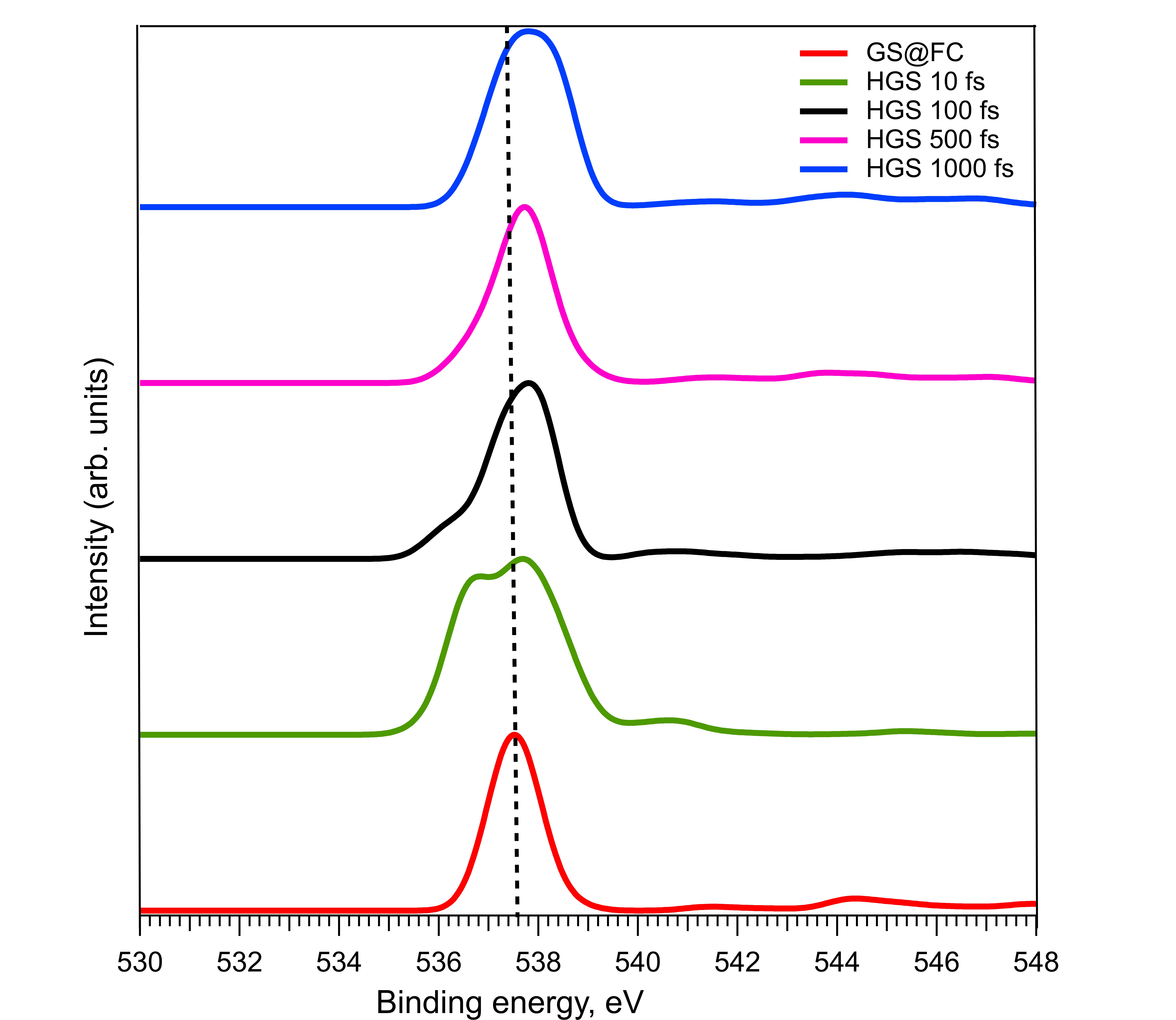}
\caption {Calculated O 1s XPS spectra for a set of geometries sampled from a ground state trajectory (not excited by the UV laser, red) and from nonadiabatic trajectories that relaxed back to the ground state (colors). The trajectories are synchronized in such a way as to start from the  $S_1(\pi\pi^*)/S_0$  CoIn at $t=0$. 
The O 1s XPS spectra are computed  at 10 fs (green), 100 fs (black), 500 fs (purple) and 1 ps (blue). All spectra are shifted by 2.4 eV to lower binding energy. The vertical black dashed line indicates the energy of the experimental GS spectrum  at 537.6 eV. \cite{feyer2009tautomerism}}
\label{fig:FIG.13} 
\end{figure}

To compute the ground state bleach signal we have performed MP2-based dynamics simulations in the electronic ground state using the same initial conditions as in SH simulations ($t=0$). 
The simulations provide the time evolution of the reference ground-state thermal ensemble. 
In addition, a second set of simulations was performed for the subset of trajectories that during SH simulations ended in the ground state giving rise to the HGS signal see Fig.~\ref{fig:FIG.13}. 
To simulate the time evolution of this HGS ensemble, initial conditions were obtained from the final ($S_1(\pi\pi^*)$/$S_0$ CoIn) geometries and velocities in the SH simulations.
 
\cleardoublepage
\section{~Calculated O 1s difference spectra.}
\label{section:7}
\begin{figure}[htb]
\includegraphics[width = 6.5 in]{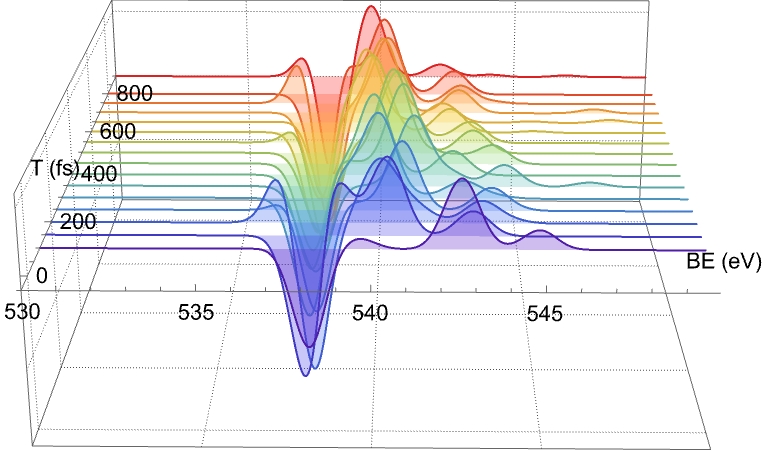}
\caption{Calculated O 1s difference spectra (excited state minus ground state) for a selected SH trajectory of uracil undergoing  internal conversion to the dark $S_1(n\pi^*)$ state. The spectra were calculated from 100 to 900 fs in 50 fs steps.}
\label{fig:FIG.14} 
\end{figure}

Figure~\ref{fig:FIG.14} shows the shake-up signal above the 546 eV region in the spectra for a selected SH trajectory which undergoes internal conversion to $S_1(n\pi^*)$ state.

.

\cleardoublepage
\section{~Depletion of the N 1s signal.}
\label{section:8}
\begin{figure}[htb]
\includegraphics[width =5.5 in]{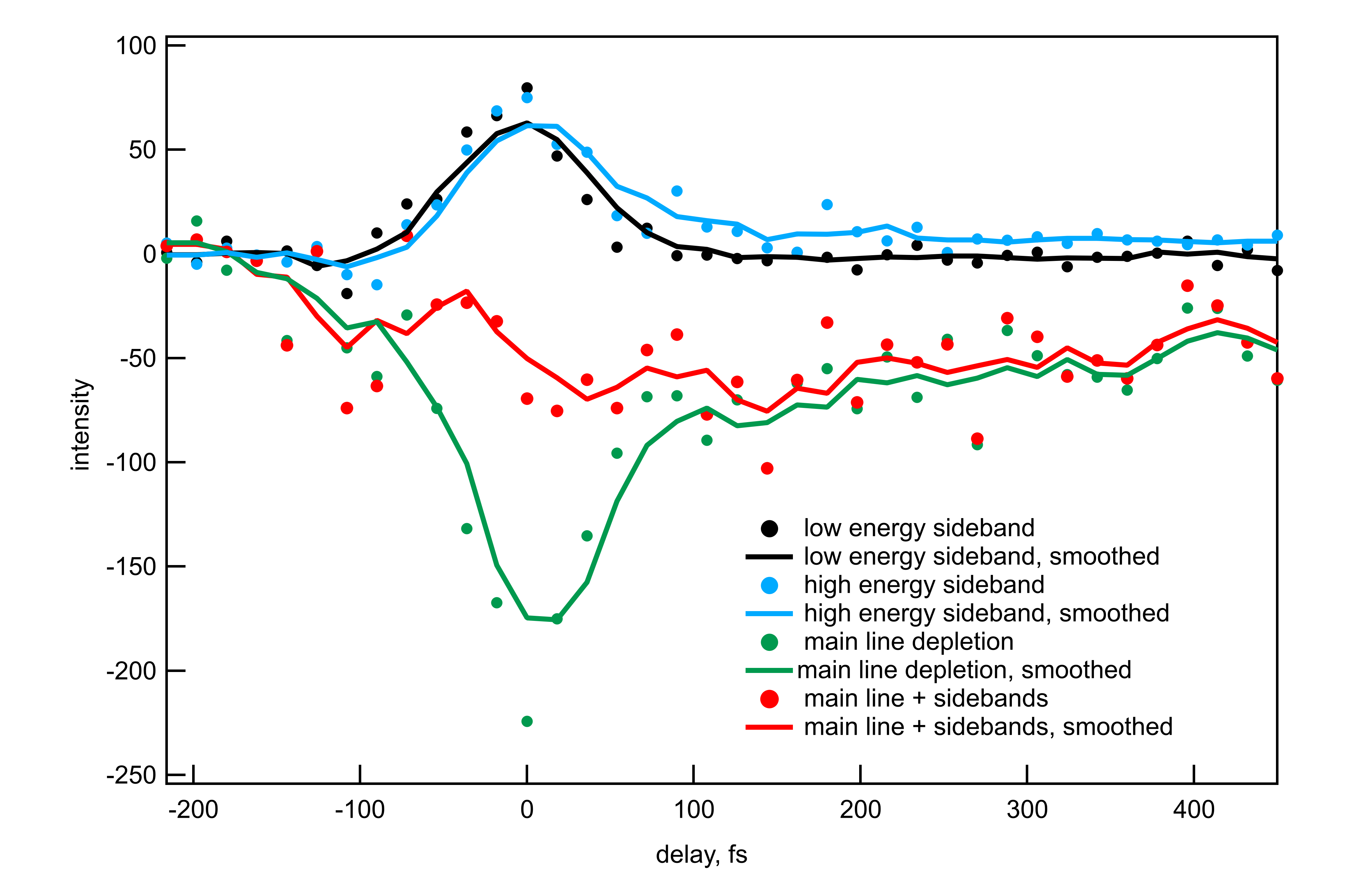}
\caption{Intensities of the N 1s features in Fig. 3(b) (main text), integrated over binding energy. 
Top curves (blue, black): lower and higher energy sidebands. 
Bottom curve (green): depletion of main line. Intermediate curve (red): sideband-corrected depletion of the main line (sum of sideband and depletion curves.)} 
\label{fig:FIG.15} 
\end{figure}

Figure~\ref{fig:FIG.15} shows the intensity of the sidebands of the N 1s signal and of the depletion of the main line, integrated over the binding energies of the respective peaks. It can be seen that adding the intensity of the two sidebands to the depletion signal cancels out the anomalous recovery of the depletion.

\cleardoublepage
\section{~Time dependence of potential energies.}
\label{section:9}
\begin{figure} [htb]
\begin{center}
\includegraphics[scale=0.3]{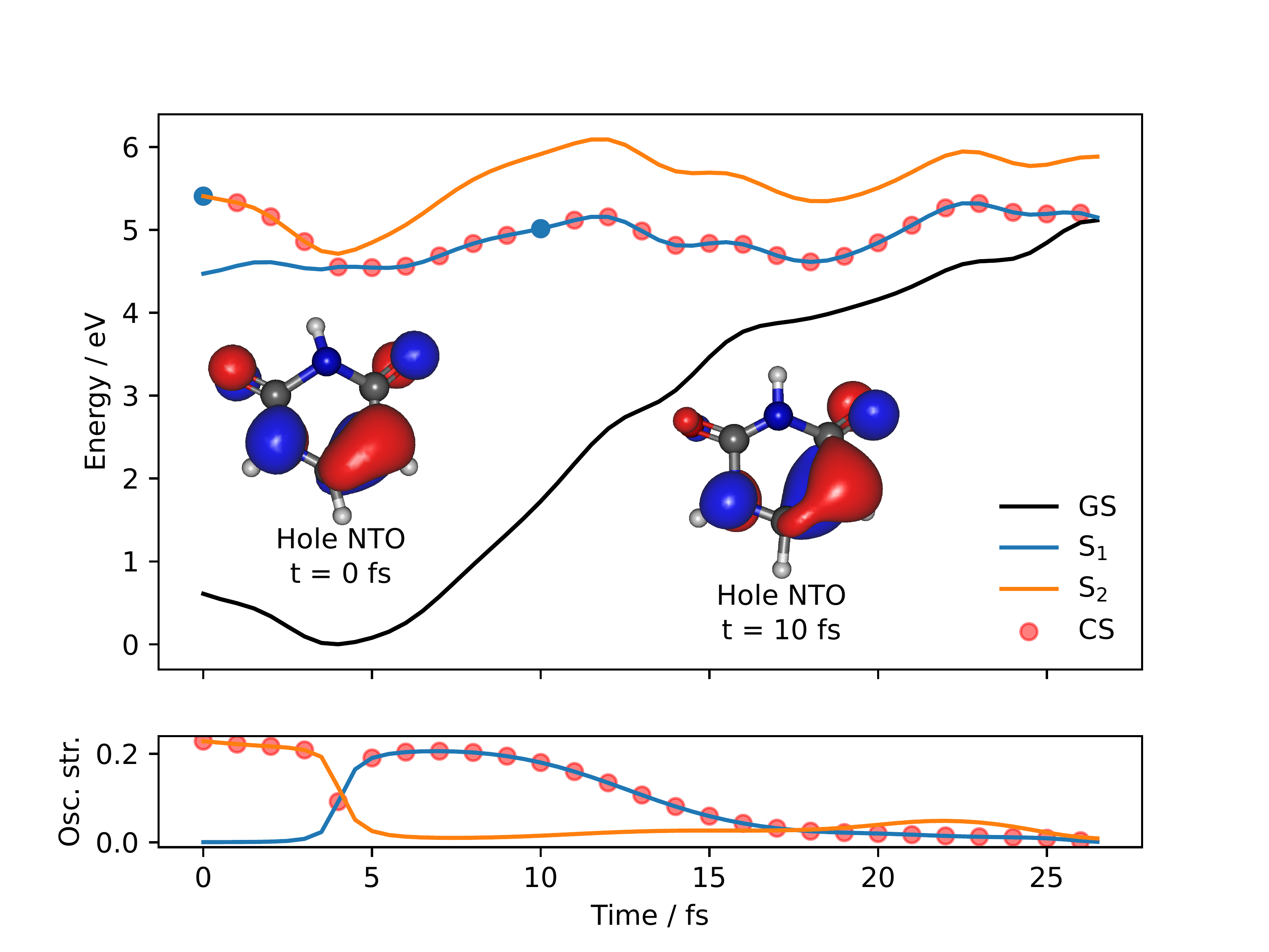}
\caption{Top: time evolution of the potential energies of the two lowest singlet excited states. 
Orange: $S_2$. Blue: $S_1$. 
Black: ground state, $S_0$. 
The pink dots indicate the populated current state (CS) at a given time. 
The time steps for the hole NTOs (colored as blue dots) are also depicted in order to indicate that the currently populated states remain the same and do not change to the '\textit{n}' orbital. 
Bottom: the oscillator strengths of the $S_2$ and $S_1$ states. 
The CoIn between the initially populated  $S_2(\pi\pi^*)$ and the $S_1(n\pi^*)$ states is encountered at around 5 fs.
Analysis of the time evolution of the oscillator strength indicates that at the  $S_2(\pi\pi^*)/S_1(n\pi^*)$  CoIn the system remains in the diabatic  $\pi\pi^*$ state. 
The  CoIn between the $S_1(\pi\pi^*)$/$S_0$ states is reached at around 25 fs.} 
\label{fig:FIG.16}
\end{center}
\end{figure}

Figure~\ref{fig:FIG.16} demonstrates that the trajectory remains in the diabatic $\pi\pi^*$ state and undergoes direct deactivation $S_2(\pi\pi^*)\rightarrow S_1(\pi\pi^*)\rightarrow S_0$  by internal conversion to the ground state {\it{via}} the ethylenic type $S_1$/$S_0$ CoIn. 

\cleardoublepage
\section{~~Calculated average bond lengths.}
\label{section:10}
To investigate the origin of oscillations in the O 1s signal intensities, we first analyzed the system in terms of ground-state normal modes. After minimizing the RMSD (root-mean-square deviation) between the excited and reference ground-state geometries, we projected all geometries from all SH trajectories onto the MP2/cc-pVDZ normal mode displacement vectors. The averaged normal mode displacements suggested potential candidates for the observed low-frequency oscillations. However, the complexity of the system and the significant displacement from the ground-state equilibrium following UV excitation make a definitive assignment challenging.

\begin{figure}[htb]
\includegraphics[width = 6.5 in]{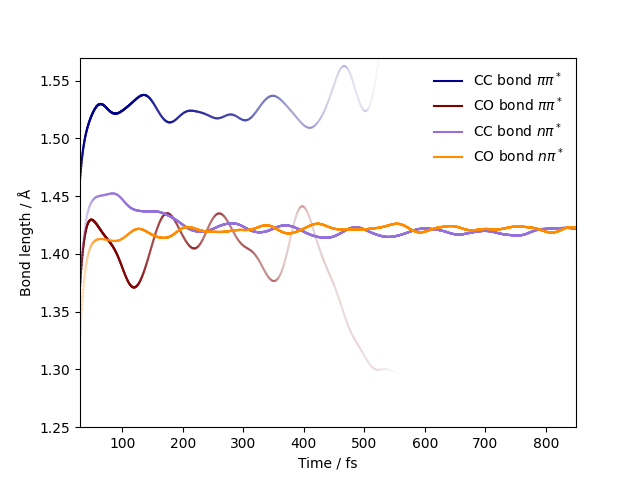}
\caption{Average bond distances of C5=C6 (CC) and C4=O8 (CO) calculated over a set of trajectories, and convoluted with a Gaussian function of 40 fs FWHM for the $(\pi\pi^*)$ and $(n\pi^*)$ states of uracil. The opacity of each line matches the number of trajectories which are in the corresponding state. 
The fading of the upper and lower lines is due to the depopulation/population of the $S_2/ S_1$  states, respectively.}
\label{fig:FIG.17} 
\end{figure}

To further explore these oscillations, we computed the average lengths of the C5=C6 and C4=O8 bonds, which exhibit substantial elongation within the first $\sim$ 20 fs of SH dynamics in the $S_2(\pi\pi^*)$ state.  Specifically, the C4=O8 bond extends from 1.26 \text{\AA} to 1.48 \text{\AA}, while the C5=C6 bond increases from 1.40 \text{\AA} to 1.63 \text{\AA}. We analyzed trajectories in the $S_2(\pi\pi^*)$ and $S_1(n\pi^*)$
states separately. As shown in Fig.\ref{fig:FIG.17}, oscillations are more pronounced in the  $\pi\pi^*$  state due to the sudden and simultaneous elongation of both bonds upon  $\pi\rightarrow \pi^*$  excitation.
Oscillations in the $n\pi^*$ state are weaker, as this state is populated more gradually at later times, allowing partial vibrational energy redistribution into other modes.
Nevertheless they remain clearly visible.  Our calculations indicate that the C5=C6 and C4=O8 bond distances oscillate with periods of approximately 100 fs and 80 fs, respectively, in good agreement with the observed signal modulations. Based on this, we interpret the periodic intensity variations in the $S_2$ and $S_1$ states as a vibronic effect.

\cleardoublepage
\section{~~Geometries of the relevant minima and conical intersections.}
\label{section:11}
\begin{figure}[htb]
\includegraphics[width =6 in]{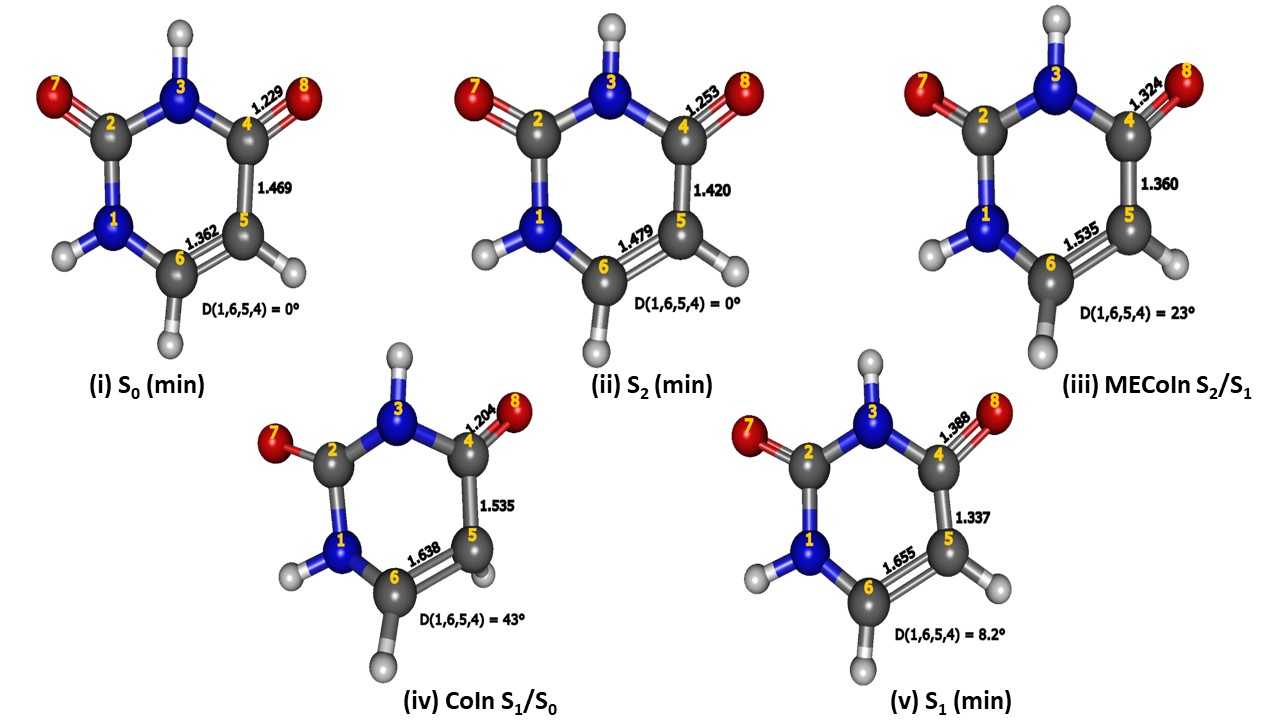}
\caption{ Relevant minima (i, ii and v) and conical intersections (iii and iv) for the photophysics of uracil. Selected bond lengths (labeled in {\AA}) are presented only where significant structural changes take place.} 
\label{fig:FIG.18} 
\end{figure}

Figure~\ref{fig:FIG.18} illustrates the five relevant structures of uracil: $S_0$ (i) and $S_2$ (ii) minima, $S_2$/$S_1$ (iii)  minimum energy CoIn (MECoIn),  $S_1$/$S_0$ (iv) CoIn, $S_1$ (v) minimum, respectively of uracil. 
The structures were optimized at the SCS-ADC(2)/aug-cc-pVDZ level of theory.
Going from the $S_0$ to the $S_1$ minimum, the main geometrical changes observed are the elongation of the C4=O8 and C5=C6 double bonds, and the simultaneous shortening of the C4-C5 bond. 
There are two important CoIns in uracil that are characterized by a strong out-of-plane distortion at the C5 carbon atom (see Figs.~\ref{fig:FIG.18}(iii) and (iv)). 
The present SH calculations indicate that the ultrafast initial change of uracil geometry (a combination of bond stretching, bending, torsion, etc.) takes place within 20-30 fs and after that, the molecule starts to vibrate around the new equilibrium structure.

\cleardoublepage
\section{~~Time-resolved C 1s spectra of uracil.}
\label{section:12}
\begin{figure}[htb]
\includegraphics[width=4.5in]{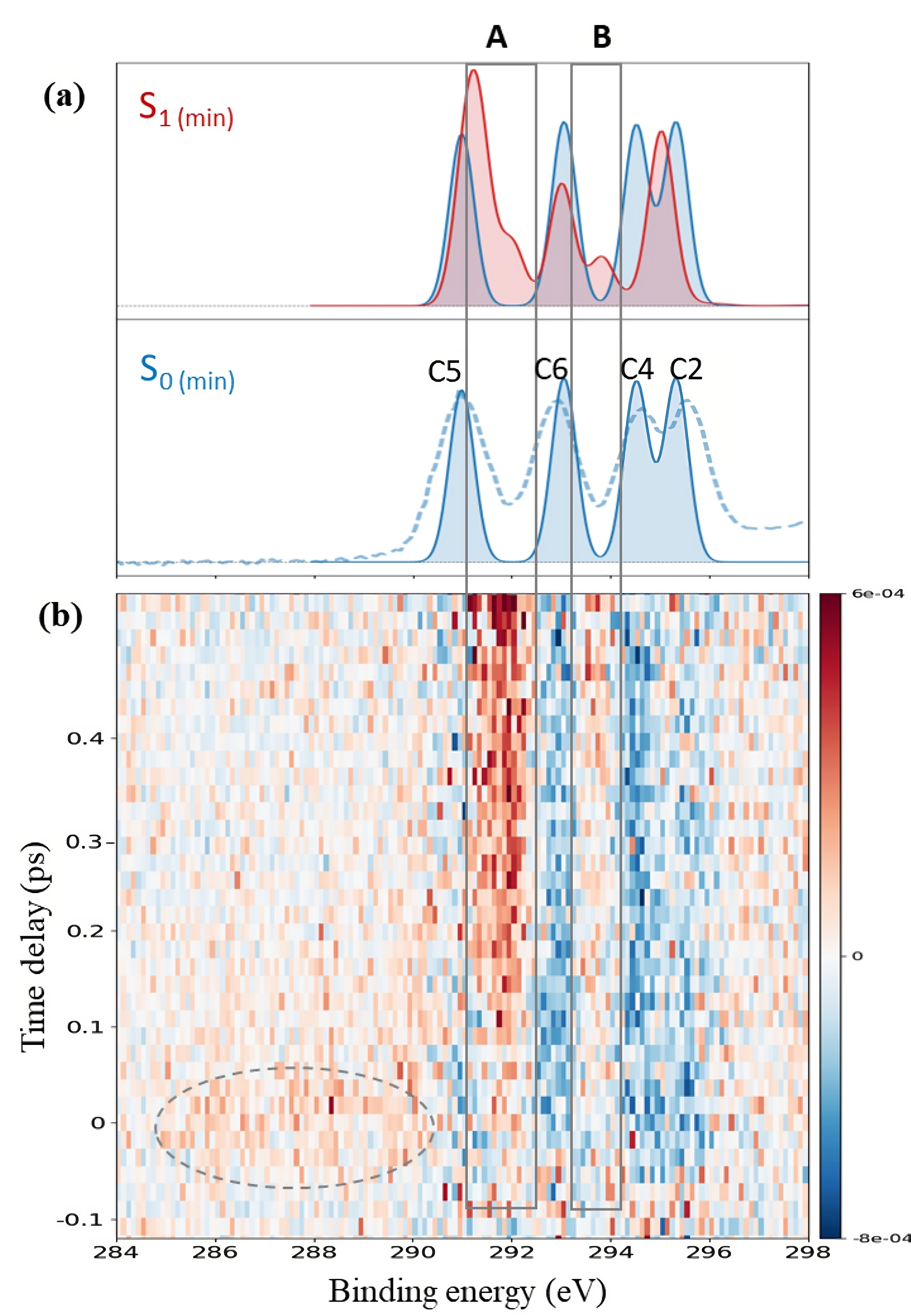}
\caption{(a) Calculated C 1s spectra of the $S_0$ ground state (blue) and the excited $S_1$ (red) state of uracil computed at their minimum energy geometries. 
Theoretical spectra are shifted by 2 eV to lower binding energy. 
The blue dashed line represents the experimentally measured GS spectrum of uracil. 
(b) Two-dimensional false color map of the C 1s subtraction spectra (UV-on minus UV-off) as a function of binding energy and of time delay (red: positive signal, blue: negative signal). 
Gray dashed circle indicates four sidebands (see text). Energy ranges (eV): (A) 291.0 - 292.5 and (B) 293.0 - 294.3.} 
\label{fig:FIG.19} 
\end{figure}

The theoretical and experimental C K-edge time-resolved photoelectron spectra of UV-excited uracil are presented in Figure \ref{fig:FIG.19}. 
Calculated C 1s spectra (see Fig.~\ref{fig:FIG.19}(a)) of the electronic ground ($S_0$) and the lowest energy excited ($S_1$) states were computed for the appropriate minimum energy structure geometries (see Fig.~\ref{fig:FIG.18}(i and v)). 
 
Uracil has four carbon atoms in its structure, hence the observed negative signal along the whole time window at 291 eV, 292.8 eV, 294.4 eV, and 295.4 eV is assigned to the ground state depletion of C5, C6, C4 and C2, respectively \cite{feyer2009tautomerism} (see Fig.~\ref{fig:FIG.19}(b)). 
The intensity at low BE (from 285 to 290 eV) around $t = 0$ is due to the four sideband signals, and is consistent with the number of carbon atoms.
 
The negative charge accumulation on the C6, C5 and C4 atoms is responsible for the initial shift for these carbon atoms relative to the GS spectrum immediately after  photoexcitation (see Table \ref{tab:charges}).
However, the C 1s TR-XPS spectra measured here turned out to be more sensitive to the structural deformation than to the partial charges. 
According to Matsika et al.\cite{matsika2004radiationless} internal conversion from $S_2\to S_1$ occurs through the conical intersection caused by bond elongation in a planar geometry (see Fig.~\ref{fig:FIG.18}(iii)). 
On the other hand, the sub-30 fs relaxation pathway from $S_2\to S_0$ is induced by a twist of the ethylenic C5=C6 double bond and a strong out-of-plane distortion of the uracil ring  (see Fig.~\ref{fig:FIG.18} (iv)) \cite{milovanovic2021simulation,matsika2004radiationless,carbonniere2015intramolecular,richter2014ultrafast,miura2023formation}.
Hence, the experimentally observed and computed spectrum for the $S_1$ state asymmetric signal at around 292 eV (see Fig.~\ref{fig:FIG.19}(a, b), range A) is assigned to the different deactivation channels (direct and indirect) induced by deformations in the uracil structure, with the biggest changes happening in the vicinity of the carbon C5 atom. 

Note that C K-edge simulations as a function of time were outside the scope of the present work due to the limited statistics and the higher computational cost compared to the O and N K-edges.

\cleardoublepage
\bibliography{References1}